\documentstyle[art12,bezier,emlines2]{article}
\normalbaselines
\input{tcilatex}
\begin{document}
\voffset=-3cm
\begin{center}
{\LARGE {\bf Physical sense of renormalizability}}\\[5mm]
{\bf B. P. Kosyakov}\\[5mm]
{\sl Russian Federal Nuclear Center -- VNIIEF, Sarov, 607190, Russia}\\[2mm]
{E-mail address: \verb|Kosyakov@vniief.ru|}
\end{center}
\begin{abstract}
A plausible physical interpretation of the renormalizability condition is
given. It is shown that renormalizable quantum field theories describe such
systems wherein the tendency to collapse associated with vacuum fluctuations
of attractive forces is suppressed by vacuum fluctuations of kinetic energy. 
Relaying on the classification of topological types of evolution of
point particles and analysing the problem of the fall to the centre, we
obtain a general criterion for preventability of collapse which states that
the spectrum of the Hamiltonian must be bounded from below. The holographic
principle is used to explain the origin of anomalies and make precise the
relation between the renormalizability and the reversibility.
\end{abstract}

\section{Introduction}
\label{Intr} A stirring event in the modern history of high energy
physics was the awarding of the 1999 Nobel Prize in physics to Gerardus 't
Hooft and Martinus Veltman ``for elucidating the quantum structure of
electroweak interactions in physics''. Experts are well aware of their
particular credit for the proof of the renormalizability of the
Yang--Mills--Higgs model which underlies this official formulation of the
Royal Swedish Academy of Sciences. As for the general physical audience, the
importance of this advance is unlikely to be fully realizable because the
physical sense of the renormalizability has been only slightly or not at all
clarified in the literature. It is felt, there comes a time where the
discussion of the renormalizability on a level close to intuitive is quite
relevant.

One should distinctly discriminate between the notion of the {\it %
renormalization} and that of the {\it renormalizability}. Our aim is to take
a close look at the latter; the physical sense of the former is sufficiently
clear, so we restrict ourselves to the following brief remark.

It was known even in the 19th century that the renormalization of mass is to
be expected in systems with infinite degrees of freedom. Let us imagine a
spherical body of mass $M_{0}$ moving with speed ${\bf v}$ through a fluid.
The hydrodynamics states that the kinetic energy of the system composed of
this moving body and a portion of the fluid dragged by it is ${\scriptstyle%
\frac{1}{2}}\,M{\bf v}^{2}$ where $M=M_{0}+\delta M$, with $\delta M$ being
the apparent additional mass equal to half the mass of the fluid displaced
by the body, and, besides, a force applied to the body produces the 
acceleration inversely related to $M$. Thus dynamical laws in fluids remain 
identical to those in vacuum, but the mass appearing in these laws can vary
considerably. For example, let a ping-pong ball be placed in a vessel
filled with mercury, then we are dealing with an object which is 80 times as
massive as that is in the habit. If the body cannot be withdrawn from the
fluid, its inertia is specified by the renormalized mass $M$, and $M_{0}$
cannot be measured. Reasoning from the analogy between the hydrodynamical
medium and the ether, J. J. Thomson introduced the electromagnetic mass of a
charged particle $\delta m$ which must be added to its mechanical mass $m_{0}
$ to give the observable mass $m$. ({\small For further elaborations of this
idea by Thomson, Lorentz and Kramers see \cite{Dresden}}.)

In the relativistic quantum domain, we encounter processes of creations and 
annihi\-la\-tions of particles. They give rise to a specific quantum field
phenomenon, the {\it vacuum polarization}, which is responsible for the
coupling constant renormalization. Indeed, let some electron be localized
within a region of size less than half its Compton wavelength. Then, due to 
the Heisenberg uncertainty principle, this will result in the energy
fluctuation sufficient for the creation of a virtual pair of an electron and a
positron. The less the localization region, the greater are the energy
fluctuations, and hence the greater is the number of pairs that might be
created and annihilated here. The electron attracts virtual positrons and
repulses virtual electrons, thus dipoles of the separated pairs screen its
charge. The electron charge $e$ measured at a distance greater than its
Compton wavelength is renormalized relative to its bare charge $e_{0}$ owing
to this screening. Furthermore, the mass $m$ of the electron wrapped up in
the coat of virtual pairs turns out to be renormalized relative to its
bare mass $m_{0}$.

Thus the vacuum of the quantum field theory (QFT) plays the role of a medium
that renormalizes masses and coupling constants. The trouble, however, is that 
such renormalizations are infinite for physically interesting Lagrangians, or, 
more technically, the calculation of corrections to masses and coupling
constants following the Feynman rules is confronted by ultraviolet
divergences.

Mathematically, the presence of these divergences is due to the fact, first
established by N. N. Bogoliubov \cite{Bogoliubov}, that the multiplication
of distributions is ill defined. The renormalization theory enabling the
absorption of ultraviolet divergences by infinite renormalizations of masses
and coupling constants gives an unambiguous prescription for the definition
of the product of propagators at points where their arguments coincide. But
this prescription is far from all-inclusive. According to the
renormalization theory, all the local quantum field theories can be
separated into two classes, renormalizable and nonrenormalizable. Take for
example a system specified by fields with spins 0 and 1/2\footnote{%
Throughout this paper we choose, unless otherwise indicated, units 
such that $\hbar =c=1$.}.
Let the Lagrangian of interaction ${\cal L}_{I}$ be a polynomial in fields,
with the monomial of the power $i$ containing a product of $b_{i}$ scalar
and $f_{i}$ Dirac fields, and $k_{i}$ derivatives. The rule of the
renormalizability by the power-counting, or, more precisely, by the index 
\begin{equation}
\omega _{i}=b_{i}+{\scriptstyle\frac{3}{2}}\,f_{i}+k_{i}-4  \label{omega_n}
\end{equation}
%(omega_n)
reads: The theory is {\it superrenormalizable} when $\omega _{i}<0$ for all $%
i$, {\it renormalizable} when $\omega _{i}\leq 0$, and {\it nonrenormalizable} 
when $\omega _{i}>0$ if only for a single $i$\footnote{This rule can be 
reformulated as follows: The theory is renormalizable when
the field dimensionality of ${\cal L}_{I}$
expressed through the dimensionality of mass $\mu $ is $\mu ^{4+\omega }$ with 
$\omega \leq 0$, and hence every coupling constant is either dimensionless
or of dimension of mass to a positive power.}.

At present the renormalization procedure, known in the mathematical
literatu\-re as the $R$-operation, is developed in every respect with a
suitable rigor ({\small the way of its development and current status are
reviewed in \cite{Zavialov4}}). There are textbooks where this procedure is
presented rather skilfully, e. g., \cite{bogshirk}--\cite{Collins}. The
study of its applied aspects, in particular the calculation of multiloop
diagrams in renormalizable theories, continues on its way \cite
{Zavialov4,Tkachov}. The mathematical nature of the $R$-operation gradually
deepens; it became clear recently that it is a special instance of a general
mathematical procedure of multiplicative extraction of finite values based
on the Riemann--Hilbert problem \cite{Connes}.

As to the notion of the renormalizability, it went a long but still
incomplete way of development ({\small impressed in excellent surveys, e.g., 
\cite{WeinbergNob}--\cite{Gell-Mann}, and books \cite{Brown,Schweber}}).
Referring the reader for details and the bibliography to these and quoted
below works, we recall only some facts directly related to our subject. Of
course, our short excursion into the history of the renormalizability is on
no account intended to the role of an express-analysis of key events in the
QFT during the last 50 years, and mentioning selected names and papers does
not imply our priority preference or evaluation of importance of any
advances. We would like only to clear up motivations of `old' and `new'
activities that, in one way or another, have a bearing on the
renormalizability.

In 1949, F. Dyson \cite{Dyson} showed that the renormalization of masses and
charges is sufficient for the removal of ultraviolet divergences in quantum
electrodynamics (QED). He observed also that it is possible to absorb all
the infinities by a redefinition of a finite number of parameters in the
Lagrangian only for a certain kind of theories that he called
renormalizable. Since then the renormalizability became a criterion for the
theory selection. In the 1970s, this criterion was slightly revised: A
theory is taken to be consistent when it is not only renormalizable by the
power-counting rule but also free of anomalies of local symmetries (even
though anomalies of global symmetries are harmless and even desirable). The
divergence phobia was replaced by the interest in such an infrequent
occurrence when they can be removed: ``The divergences of quantum field
theory must not be viewed as unmitigated defects; on the contrary, they
convey crucially important information about the physical situation, without
which most of our theories would not be physically acceptable \lbrack
...\rbrack\ One cannot escape the conclusion that Nature makes use of
anomalous symmetry breaking, which occurs in local field theory owing to
underlying infinities in the mathematical description'' \cite{Jackiw}.

We draw attention to the instrumental character of the
criterion. It reflects the view on the acceptable QFT as a collection of
perturbation rules compatible with the renormalization procedure.
Nonrenormalizable theories are thought to be bad not due to the fact that
they are based on some physical `pathologies' but merely because we do not
know how to handle them. All reverses of fortune of the criterion are
related to this feature. At different times, the perception of the criterion
changed drastically, from decisive rejection of all local theories, both
renormalizable and nonrenormalizable, to full tolerance of every theory,
nonrenormalizable including; yet one attempted seldom if ever to clarify the
difference of renormalizable and nonrenormalizable theories in their
physical essence.

The crisis of the 1950--1960s in QFT burst out in connection with the
discovery of the `null-charge' phenomenon by L. D. Landau and I. Ya.
Pomeranchuk \cite{LandauPomeranchuk}, and independently by E. S. Fradkin 
\cite{Fradkin}. They revealed that the vacuum polarization in QED and other
models is so strong that the observed coupling constants are subject to a
complete screening irrespective of values of bare coupling constants.
Another troublesome surprise was the discovery of the photon `ghost' state 
\cite{LAKh}. Landau regarded the charge nullification as evidence of logical
inconsistency of QED and the conception of local interactions as such \cite
{Landau}. The renormalizability principle was buried under debris of the
Lagrangian formalism for a long time. ``Under the influence of Landau and
Pomeranchuk, a generation of physicists was forbidden to work on field
theory'' \cite{Gross}.

In the early 1970s, the `gauge revolution' occured and the `Golden Age' of
renorma\-lizab\-le theories came. The breakthrough rested on the proof of the
renormalizability of the Yang--Mills--Higgs model \cite{Hooft0} and the
discovery of the asymptotic freedom \cite{GrossWilczek}. The renormalizable $%
SU(3)\times SU(2)\times U(1)$ gauge theory of strong, electromagnetic and
weak interactions was established and received the name Standard Model. The
idea of a confluence of running coupling constants in the vicinity of $%
10^{16}$ GeV initiated efforts to build the Grand Unification of three
fundamental interaction in the framework of gauge theories with a simple
group of internal symmetry, for instance $SU(5)$ or $O(10)$. All these
milestones are now widely known and are expounded in textbooks.

It was anticipated that the availability of the asymptotic freedom would
rehabilitate the four-dimensional (4D) quantum field theory. ``One can trust
renormalization theory for an asympto\-tically free theory, independent of
the fact that perturbation theory is only an asymptotic expansion, since it
gets better and better in the regime of short distances'' \cite{Gross}. One
believed that the $S$ matrix in such theories would be free of the Landau
ghost. However, it became soon clear that the ghost simply migrated from the
ultraviolet region to the infrared. As to the quantum chromodynamics, this
finding implied only that the confinement is a nonperturbative effect.
Nevertheless, one was forced to bid farewell to the dream that asymptotic 
theories are consistent already on the perturbation level.

The proximity of the Grand Unification scale $10^{16}$ GeV to the Planck
mass $M_{{\rm P}}= k^{-1/2}=1.22\times 10^{19}$ GeV where $k$ is the Newton
gravitation constant was an impetus to the unification of all four
fundamental forces, gravity including. But a short time later, the
renormalization ideology reached a deadlock for the nonrenormalizability of
gravitation.

By the end of the 1970s, rapidly grew up the supergravity \cite
{Nieuwenhuizen}. Due to a remarkable feature of supersymmetric Yang--Mills
theories, the cancellation of divergences in the one-loop approximation
(some theories prove finite in all perturbation orders), a cardinal change
of attitude had been seen on what ultraviolet behavior should be demanded
from consistent field theories. The requirement of  renormalizability was
replaced by the condition of  {\it finiteness} \cite{Duff2}. Notice, we are
concerned not with a cutoff at the Planck length $l_{{\rm P}}=1.6\times
10^{-33}$ cm, but with a cancellation of divergences which must be ensured
by the `true' field contents. One looked to the eleven-dimensional ${\cal N}%
=1$ supergravity since 11 is both the minimal dimension for $SU(3)\times
SU(2)\times U(1)$ to be the internal symmetry group and the maximal
dimension compatible with the supersymmetry of fields with spin $J\leq 2$. A
mechanism of compactification of 7 redundant dimensions  was invented 
in the spirit of the Kaluza--Klein ideology.

However, the project failed: The 4D on-shell supergravity turns out to be
finite only up to the two-loop approximation while the 11D supergravity
suffers even from one-loop divergences. Furthermore, any odd-dimensional
theory cannot be chiral, and the compactification gave no way
of deriving a chiral 4D theory describing the observed world with the
inherent asymmetry between the right and the left from the 11D supergravity.

Theorists fastened their eyes on superstrings \cite{GSW}--\cite{Polchinski99}. 
The quantum string theory is in general free of ultraviolet divergences
but is plagued by anomalies. Five superstrings was shown to be free of
anomalies: The open string with the ${\cal N}=1$ supersymmetry and the $%
SO(32)$ gauge symmetry, the type I string, two closed strings with the $%
{\cal N}=2$ supersymmetry, chiral and non-chiral, the type IIA and IIB
strings, and two closed strings with different constructions of right and
left sectors possessing the $E_{8}\times E_{8}$ and $SO(32)$ gauge
symmetries, the heterotic strings. A consistent quantization of superstrings
is feasible only in 10 dimensions. The most promising was the heterotic $%
E_{8}\times E_{8}$ string since it can incorporate the Standard Model. It
remained only to compactify spacetime from 10 to 4 dimensions and reduce too
high supersymmetries. The Calabi--Yau manifolds and orbifolds offered
prospects for tackling these tasks.  A consistent theory containing both
semiclassical supergravity and Grand Unification with the chiral representation
of quarks and leptons in the low energy limit eventually resulted. It was
named  `Theory of Everything'.

Why strings are free of the ultraviolet diseases? The supersymmetric
cancellation is `almost immaterial'; bosonic strings are finite as
well, but there is a tachyon in their spectrum which is removable with the aid
of the supersymmetry. The finiteness of the string theory is often
associated with the fact that strings are nonlocal. Strings are extended
objects, and this should presumably provide the cutoff near the energy scale $%
M_{{\rm P}}$. But the string interaction is known to be local, for example,
two open strings merge into a single one only when their ends
contact. Meanwhile the cutoff due to nonlocal form-factors means that the 
{\it interaction is smeared out} over a finite region \cite{Efimov}. If the 
interaction is not smeared out, then, as was shown in \cite{Trivedi}, 
divergences do not disappear\footnote{%
We will return to the discussion of this instructive result at the end of
the paper.}. The reason for the good ultraviolet behavior of strings is
likely to be the fact that we are dealing with a local conformal QFT on 
{\it two-dimensional} manifolds that are formed by the world sheets swept out
in the course of the string evolution. In some cases this two-dimensional
conformal theory is actually the only which is at our disposal. For example,
in the heterotic construction, $X_{\mu }$ is a boson field that cannot be
interpreted as the string coordinate in the target spacetime. That the
finite length of strings does not assure the ultraviolet finiteness 
can be seen from the comparison of strings with membranes. Indeed, attempts to
build an appropriate local QFT on three-dimensional world volumes of the 
membrane revert us to the problem of ultraviolet divergences.

So, the string theory is free of ultraviolet divergences. But in lieu of
them, new conceptual difficulties arose. Most notable of them is the
uniqueness problem. The Theory of Everything must be unique by its very
nature. It should stand out against all possible theoretical schemes not
only because of its best phenomenology reflection but also because of the
lack of mathematical inconsistencies, unique to it. We have instead
five consistent theories. Things get worse if we compactify 6 redundant
dimensions: An abundance of the Calabi--Yau configurations was found
which describe vacuum states of the same energy, and it is not clear how this
collection can be narrowed.

Another way of solving this problem is to show that all these theories are
the manifestation of the same physics, but in various contexts, e. g., in
regimes of strong and weak coupling, i. e., every two of them are linked by
some duality transformation \cite{Polchinski96}. Lately, it was understood
that this is actually the case: All superstrings and soliton-like objects of
the 11D supergravity (branes, black holes, etc.) are tightly entangled in a
duality web as a unified scheme, that was called the M theory \cite{Witten95}. 
(A rigorous justification of the dualities would be possible if we were
aware of all nonperturbative solutions of the theories under examination.
Unfortunately, one failed to find exact solutions in the strong coupling
regime, and, therefore, many dualities remain plausible conjectures.) A
striking feature of the M theory is that it describes some 11D realm, with
the chiral $E_{8}\times E_{8}$ string emerging due to the compactification
realized on a line-segment. Another property is the presence of extended
objects of different dimensions capable to mutual conversions. At
present, it is not clear what degrees of freedom are fundamental in M
theory, in any case, those are not strings or branes ({\small for a readable
review of M theory see \cite{Duff8}}).

This opens the door to construct the finite, free of anomalies, unique 
theory\footnote{Meanwhile it should be remembered alternative approaches, 
e.g., the loop quantum gravity \cite{Rovelli}.} embo\-dy\-ing the first 
principles of the physical world.
Is renormalizability as yet required as a fundamental principle? Maybe
it is an appropriate time to say good-bye to it. One has at least two
objection to this. Firstly, the M theory is a formidable project. It is 
unlikely that success will be achieved if one takes no lesson from 
pre-stringy efforts to cope with the treatment of systems with infinite
degrees of freedom. (Were these efforts actually doomed to failure?) Secondly,
as R. Jackiw \cite{Jackiw} noted: ``I wonder where within the completely
finite and non-local dynamics of string theory are we to find the mechanisms
for symmetry breaking that are needed in order to explain the world around
us''. Indeed, how can we reasonably explain the symmetry violations
differently than considering them as healed up scars of ultraviolet
wounds? In the flawed symmety problem, the finitism cease to be the good.

Besides, there is a more technical problem. Lowering ourselves from the
Planck height into the region of experimentally attainable energies, we
cannot take advantage of the finiteness of the string theory for
calculating processes with usual particles, for example quarks, since we do
not know the genuine mechanism of the dimension reduction from 10 to 4. For
comparison recall that classical relativistic mechanics makes it possible
to calculate both relativistic and non-relativistic motions; in some cases
(for example, the particle motion in a plane electromagnetic wave) this
calculation turns out to be even more simple than that in Newtonian
mechanics. Thus the string theory gave no answer to the question put in the
title of the famous Gell-Mann talk at the Shelter Island Conference in 1983 
\cite{Gell-Mann}.

Let us turn to last pages of the history of the renormalizability ({\small %
for more extended reviews see \cite{Cao,SchweberR}}) and raise the naive
question: Why are three fundamental interactions, strong, electromagnetic,
and weak, renormalizable? It would be ridiculous to think that Nature plays
at give-away with us arranging part of laws in such a way that we were able
to realize them by means of perturbative calculations.

In the 1970s, approaches levelling differences between renormalizable and
nonrenorma\-liz\-able theories came into particular prominence. K. Wilson
revived interest in the renormalization group as a useful tool in studying
the situation in QFT beyond the perturbation theory ({\small for the history
and present status of the renormalization group investigati\-ons see \cite
{Shirkov}}). By analogy with the critical point in phase transitions of the
second kind, where fluctuations of all scales are comparable in magnitude,
one can define the notion of ultraviolet fixed point, approaching to
which renders quantum fields asymptotically scale invariant ({\small for
more detail see \cite{Wilson}}). This notion is the starting point in the S.
Weinberg program of `asymptotic safety' which states that coupling
parameters tend to a fixed point when the scale of their energy
normalization tends to infinity \cite{Weinberg79}. For the Gaussian fixed
points, the asymptotic safety is equivalent to the renormalizability in the
usual sense (with the null-charge problem being conclusively avoided), while,
in the general case, it was intended to enlarge the renormalizability
condition. But still no one was able to prove or disprove the availability 
of fixed points in realistic 4D gauge theories such as QED.

By the mid-1970s, some progress in the constructive QFT occured that sheded
light on the structure of nonrenormalizable theories and conditions under
which these theories acquire the mathematical sense. In particular, it was
found that some nonperturbative solutions of such theories can be specified
by a finite number of arbitrary parameters as opposed to the situation in
the perturbation theory ({\small for a review of this progress see \cite
{Wightman}}). At present, the idea to weaken the renormalizability requirement 
and extend the class of physically admissible theories has lost its acuteness, 
and yet it does not consign to oblivion \cite{GomisWeinberg}.

Another line of the renormalization group investigations culminated in
the proof of the `decoupling theorem' according to which, in a renormalizable
theory with a field of mass $M$ much greater than masses of other fields, it is
possible to find such renormalization prescription that the heavy field
decouples and reveal itself only as corrections to the Lagrangian suppressed
by powers $E/M$ where $E$ is an energy characteristic to this region (%
{\small for detail see chapter 8 of the book \cite{Collins}}). An important
consequence is that low-energy physics is described by effective theories
containing only those particles which are actually significant in the energy
range under consideration. Thus the mass $M$ plays the role of ultraviolet
cutoff in the effective theory; if $E\rightarrow M$ then the effective
theory fails to be applicable and should be replaced by a new effective
theory with greater cutoff. For $E\ll M$, the heavy particles can reveal
themselves only through the processes (for example, weak decays) which are
forbidden by symmetries (for example, by the parity conservation) in the
absence of the heavy particles (for example, $W$ bosons) which mediate the
interaction with the broken symmetry. These small effects correspond to
nonrenormalizable terms of the Lagrangian since they are multiplied by
inverse powers of $M$, and hence have the operator dimension of mass to a
positive power. ``Thus the only interactions that we can detect at ordinary
energies are those that are renormalizable in the usual sense, plus any
nonrenormalizable interactions that produce effects which although tiny, are
somehow exotic enough to be seen'' \cite{WeinbergNob}. For example, the
nonrenormalizable four-fermion interaction ${\scriptstyle\frac{G}{\sqrt{2}}}%
\,J\cdot J$ is suppressed by the smallness of the Fermi constant $G$
(proportional to $M_{W}^{-2}$) for $E\ll M_{W}$, and yet observable due to
the chiralness of weak processes.

What is the explanation of the renormalizability of the Standard Model in
the context of effective theories? The answer is rather evident. The
Standard Model is an effective low-energy theory that might be derived from
a higher-level theory, e. g., from the Grand Unification, if one would
integrate out all the heavy fields, e. g., $X$ bosons, in the path integral.
Since we are keeping in mind the observed world, the Lagrangian that
governs it should not be suppressed by powers of $1/M_X$, whence, 
for dimensional reasons, it follows that it is renormalizable.

With this in mind, the gravitation Lagrangian,
nonrenormalizable by the power-counting, should be suppressed by powers of $%
1/M_{{\rm P}}=\sqrt k$. However, it is well known that such is not the case: 
The Hilbert Lagrangian ${\sqrt{-g}}R/16\pi k$ is not divided but, quite the
reverse, multiplied 
by $M_{{\rm P}}^2$. Although the Einsteinian gravity emerges as the low-energy 
limit of the string theory, it bears no relation to effective theories.

So, in the framework of the effective theory ideology, we found only part of
the answer (not entirely convincing) to the above question. Moreover, we
remained ignorant of the {\it dynamical} reason by which
nonrenormalizable interactions are in disrepute or at least are
suppressed by small coefficients. It is even misunderstood whether the
renormalizability is an antipode of the nonrenormalizability.

Below we will discuss just these aspects of the renormalizability problem.
Attention will be centered on the quantum physics at distances very
short by macroscopical standards but much greater than the Planck length 
$l_{{\rm P}}$. We assume gravity as a quantum phenomenon to be of little
importance in this region; its impact amounts to the classical curvature of
spacetime. In Sec. 4 we will elucidate that the condition of 
renormalizability is equivalent to the condition of suppressibility of
collapse. But we will approach this conclusion deliberately step by step. In
Sec. 2 we will be concerned with topological types of evolution of point
particles which enables us to look at the collapse from topological
point of view. ({\small One often claims that classical theory has nothing
to do with the renormalizability problem altogether since there are no
processes of the creation and annihilation of particles in it, and, thus,
coupling constants do not subject to the renormalization. It is widely
believed that the classical self-energy $\delta m$ diverges differently than
the quantum self-energy $\Sigma $. This seems to rule out any linkage
between ultraviolet diseases in classical and quantum theories. Distinguished 
as those diseases may be, the criterion of
viability of both theories is the same, the preventability of collapse.
Furthermore, In Sec. 6 we will see that the powers of divergences of
corresponding quantities in dual classical and quantum pictures are in effect
coincide. Thus the analysis of ultraviolet properties of classical pictures
is of great benefit to gaining insight into what happens in much less clear
quantum pictures}.) The simplest variant of the collapse, the fall to the
centre, discussed in Sec. 3 allows to formulate the criterion of preventability
of collapse in a general form: The collapse is preventable if the spectrum
of the Hamiltonian is bounded from below. In Sec. 5 we will consider the
similarity and difference of the concepts of renormalizability, kinetic
dominance, and suppresibility of collapse. In Sec. 6 we will use the
holographic principle to explain the origin of anomalies and make precise
the relation between the renormalizability and reversibility. Section 7 sums
up our discussion.

\section{Topological types of evolution}
\label{Typ} There are two points of view on what is the basic object
in the classical field theory. One of them takes fields as fundamental, and
particles as special manifestations of the fields. The other takes particles
fundamental while fields play the role of codes conveying information of the
particle behavior. Feynman and Wheeler showed that both
standpoints are on an equal footing  in classical electrodynamics ({\small %
see, e. g., \cite{Feynman}}). Besides, it is well known that the 
quantization leads to a synthetic object, the quantized field.

It is convenient to begin with the classical picture turning to notions that 
are rather intuitive but very deep, the topological notions. We adopt the 
point of view by which particles are fundamental, and consider their evolution.

The only topological guide relevant to such a picture is the criterion of 
{\it compactness}. It discriminates between finite and infinite motions. In
other words, all systems of particles can be classed as bound and unbound.
Thus, decays and recombinations are topologically significant events,
while elastic scatterings are not.

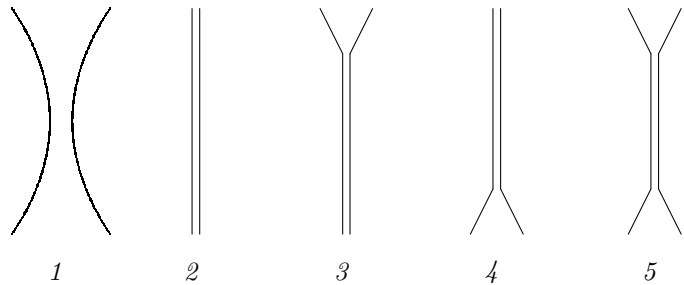
\begin{figure}[htb]
\begin{center}
\unitlength=1mm
\special{em:linewidth 0.4pt}
\linethickness{0.4pt}
\begin{picture}(95.00,40.00)
\bezier{144}(5.00,10.00)(15.00,25.00)(5.00,40.00)
\bezier{144}(18.00,40.00)(8.00,25.00)(18.00,10.00)
\emline{29.00}{10.00}{1}{29.00}{40.00}{2}
\emline{30.00}{40.00}{3}{30.00}{10.00}{4}
\emline{49.00}{10.00}{5}{49.00}{34.00}{6}
\emline{49.00}{34.00}{7}{46.00}{40.00}{8}
\emline{50.00}{10.00}{9}{50.00}{34.00}{10}
\emline{50.00}{34.00}{11}{53.00}{40.00}{12}
\emline{66.00}{10.00}{13}{69.00}{16.00}{14}
\emline{69.00}{16.00}{15}{69.00}{40.00}{16}
\emline{70.00}{40.00}{17}{70.00}{16.00}{18}
\emline{70.00}{16.00}{19}{73.00}{10.00}{20}
\emline{87.00}{10.00}{21}{90.00}{16.00}{22}
\emline{90.00}{16.00}{23}{90.00}{34.00}{24}
\emline{90.00}{34.00}{25}{87.00}{40.00}{26}
\emline{94.00}{40.00}{27}{91.00}{34.00}{28}
\emline{91.00}{34.00}{29}{91.00}{16.00}{30}
\emline{91.00}{16.00}{31}{94.00}{10.00}{32}
\put(11.00,5.00){\makebox(0,0)[cc]{${\it 1}$}}
\put(29.00,5.00){\makebox(0,0)[cc]{${\it 2}$}}
\put(49.00,5.00){\makebox(0,0)[cc]{${\it 3}$}}
\put(69.00,5.00){\makebox(0,0)[cc]{${\it 4}$}}
\put(90.00,5.00){\makebox(0,0)[cc]{${\it 5}$}}
\end{picture}
\caption{Elementary topological types of evolution}
\label
{types}
\end{center}
\end{figure}

A diagrammatic sketch of elementary topological types of evolution is
displayed in Figure \ref{types}. For simplicity, world lines of only two
particles are diagrammed, but one keeps in mind systems composed of any
number of particles. A history of an unbound system executing an infinite
motion is shown by diagram ${\it 1}$. 
A history of a stable bound system is
shown by diagram ${\it 2}$; the motions of such systems are compactly
supported. For some forms of interactions, a bound system can exist
for a half-infinite period, whereupon it decays into separate fragments, as
is shown by diagram ${\it 3}$. Here we see a topological change of the
regime of motion: From compact to noncompact. One might also conceive the
formation of a bound system existing indefinitely long (there is a finite
probability that nothing happens with it at a later time). The situation
is shown by diagram ${\it 4}$. Here we encounter the opposite regime change:
From noncompact to compact. If the formed system decays after a lapse of a
finite period, we arrive at the situation shown by diagram ${\it 5}$%
\footnote{Diagram ${\it 5}$ turns to diagram ${\it 1}$ as lifetime of the 
formed system tends to zero.}. Here we see a double topological change of the
regime of motion: From noncompact to compact, and vice versa. The collection
of variants of evolution ${\it 1}$--${\it 5}$ almost exhausts all the
possible topological types. Histories of classical systems of point
particles are built out of these elementary variants.

{\small To take an example, imagine a realm where all particles are
contained in clusters incapable of decay into individual particles, but
capable of exchanging their constituents at collisions. It is just the
situation of the cold subnuclear realm where quarks are confined in hadrons,
and do not exist in the isolated form. The bulk of the hadron phenomenology
is grasped by planar diagrams \cite{witten}, which implies in particular
that the individuality of valence quarks remains unchanged as long as they
reside in an undisturbed hadron. Such persistence of valence quarks is
peculiar to classical particles. That is why this realm can be described 
semiclassically. The planar diagrams are largely built out of the diagrams 
${\it 1}$--${\it 5}$. All one need to add to this set is the seagull-type 
diagram (two half-infinite timelike curves springing up from a single point). 
This diagram expresses the creation or the annihilation of a quark-antiquark
pair. In the classical theory, the seagull-shaped world line configurations
are forbidden due to their incompatibility with the least action principle.
Creations and annihilations of valence quarks inside hadrons are actually
suppressed by the Okubo--Zweig--Iizuka rule (see, e. g., \cite{Ogawa,Close}%
), yet a moderate amount of such processes is tolerated at hadron collisions.%
}

\begin{figure}[htb]
\begin{center}
\unitlength=1mm
\special{em:linewidth 0.4pt}
\linethickness{0.4pt}
\begin{picture}(75.00,40.00)
\bezier{136}(5.00,40.00)(13.00,25.00)(5.00,10.00)
\bezier{136}(13.00,40.00)(5.00,25.00)(13.00,10.00)
\emline{25.00}{40.00}{1}{25.00}{10.00}{2}
\emline{25.00}{10.00}{3}{25.00}{10.00}{4}
\emline{40.00}{10.00}{5}{40.00}{35.00}{6}
\emline{40.00}{35.00}{7}{35.00}{40.00}{8}
\emline{45.00}{40.00}{9}{40.00}{35.00}{10}
\emline{55.00}{40.00}{11}{55.00}{15.00}{12}
\emline{55.00}{15.00}{13}{50.00}{10.00}{14}
\emline{65.00}{10.00}{17}{70.00}{15.00}{18}
\emline{70.00}{15.00}{19}{75.00}{10.00}{20}
\emline{70.00}{15.00}{21}{70.00}{35.00}{22}
\emline{70.00}{35.00}{23}{65.00}{40.00}{24}
\emline{70.00}{35.00}{25}{75.00}{40.00}{26}
\put(9.00,5.00){\makebox(0,0)[cc]{${\widehat{\it 1}}$}}
\put(25.00,5.00){\makebox(0,0)[cc]{${\widehat{\it 2}}$}}
\put(40.00,5.00){\makebox(0,0)[cc]{${\widehat{\it 3}}$}}
\put(55.00,5.00){\makebox(0,0)[cc]{${\widehat{\it 4}}$}}
\put(70.00,5.00){\makebox(0,0)[cc]{${\widehat{\it 5}}$}}
\emline{55.00}{15.00}{27}{60.00}{10.00}{28}
\end{picture}
\caption{Special variants of evolution}
\label
{subvar}
\end{center}
\end{figure}
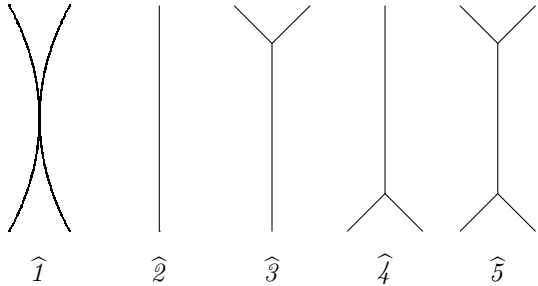

From the variants ${\it 1}$--${\it 5}$, it is possible to separate
subvariants corresponding to special states where the region of finite
motion shrinks to a single point (we recall that the set containing a single
point is compact). These special variants are displayed as diagrams similar
to diagrams ${\it 1}$--${\it 5}$, but with replacing two vertical lines by a
single one, Figure \ref{subvar}, and are numbered by hatted numbers.

The special variant of evolution ${\widehat{{\it 4}}}$ may (or may not) be
identified with the collapse. When such is the case, the special variants
differ significantly from the ordinary ones. Indeed, should the formation of
an ordinary stable bound system be allowable (the variant ${\it 4}$), in
view of reversibility, decays of it are also feasible (the variants ${\it 3}$
and ${\it 5}$). On the other hand, as will be shown in the next section, a
kind of collapse, the fall to the centre, is unavoidable, if the attraction
potential is more singular than the centrifugal term. In this case, two
particles merge into a single one, remaining amalgamated forever. Thus, the
mere presence of the variant ${\widehat{{\it 4}}}$ renders the inverse
variant ${\widehat{{\it 3}}}$ impossible (topologically, these diagrams can be
distinguished by the number of arrows flowing in and flowing out the vertex),
and also the variants ${\widehat{{\it 1}}}$ and ${\widehat{{\it 5}}}$. This
suggests a topological violation of reversibility. The picture may be
qualified as ${\widehat{{\it 1}}}$-, ${\widehat{{\it 3}}}$-, and ${\widehat{%
{\it 5}}}$-deficient.

When attraction is less singular than centrifugal effects, the fall to the
centre is prevented, and the variant ${\widehat{{\it 4}}}$ disappears. The
variants ${\widehat{{\it 1}}}$, ${\widehat{{\it 2}}}$, ${\widehat{{\it 3}}}$
and ${\widehat{{\it 5}}}$ are also missing from the picture due to lack of
the reason for the formation of merged particles. We thus arrive at a
depleted picture where all the variants ${\widehat{{\it 1}}}$--${\widehat{%
{\it 5}}}$ are absent, but the reversibility is regained. We will see
further that the separation of field theories into renormalizable and
nonrenormalizable follows from the condition for suppressibility of
collapse, which ensures the physical consistency of a given theory at the
cost of the depletion of the topological picture, that is, at the cost of \
disappearance of all the special variants.

The outlined classification may be in a one-to-one way translated into the 
{\it spectral} language\footnote{%
Notice, this language is versatile enough to get rid of reference to the
initial particle picture.}. Indeed, the variants ${\it 1}$ and ${\it 2}$
correspond, respectively, to continuous and discrete parts of spectrum. The
variant ${\it 3}$ implies the conversion of discrete spectrum of $|\,{\rm in}%
\rangle $ states to continuous spectrum of $|\,{\rm out}\rangle $ states,
the variant ${\it 4}$ refers to the opposite conversion. The variant ${\it 5}
$ is associated with a resonance line. It turns to the variant ${\it 1}$ as
life time of the bound state vanishes, and the corresponding resonance line
spreads to the extent that it ceases to be interpreted as an element of the
discrete spectrum.

{\small A less trivial illustration of this correspondence is that the
aforesaid grasping the hadronic realm by planar diagrams can supposedly be
expressed \cite{Ne'eman} in terms of Regge trajectories displayed as
straight lines with a fixed slope on the Chew--Frautschi plot of hadronic
mass squared $m^{2}$ versus spin $J$, with hadrons on any trajectory being
separated by intervals $\Delta J=2$. However, no convincing explanation of 
the correspondence between the planar diagram picture and the Regge
equidistant spectrum was still proposed.}

The special variants correspond to continuous spectra with a gap between the
vacuum and one-particle energy levels (which is larger than the gap that was
prior to the particle merger). If the variant ${\widehat{{\it 4}}}$ is
identifiable as the collapse, the depleted picture (without all the special
variants ${\widehat{{\it 1}}}$--${\widehat{{\it 5}}}$) corresponds to a
spectrum bounded from below, whereas the  ${\widehat{{\it 1}}}$-, ${\widehat{%
{\it 3}}}$- and ${\widehat{{\it 5}}}$-deficient picture bears on systems
with the energy spectrum unlimited from below.

\section{The fall to the centre}
\label{Fall} As a prelude to the discussion of the collapse in systems with
infinite number of degrees of freedom, we recall some aspects of the
relativistic Kepler problem. This is a two-particle problem reducible to the
problem of a single particle moving in a potential $U(r)$ and specified by
the Hamiltonian (see, e. g., \cite{LandauLifshitz}, Sec. 39) 
\begin{equation}
H=\sqrt{m^{2}+\frac{p_{\phi }^{2}}{r^{2}}+p_{r}^{2}}+U(r)
\label{Hamiltonian}
\end{equation}
%{Hamiltonian}
where ${p_{\phi }}$ and ${p_{r}}$ are the momenta canonically conjugate to
the polar coordinates $\phi $ and $r$. Note that ${p_{\phi }}$ is a
conserved quantity, the orbital momentum $J$. Switching off the dynamics, i.
e., taking ${p_{r}}=0$ in (\ref{Hamiltonian}), one obtains the effective
potential ${\cal U}(r)$ whereby the particle behavior near the origin is
conveniently examined, 
\begin{equation}
{\cal U}(r)=\sqrt{m^{2}+\frac{J^{2}}{r^{2}}}+U(r).  \label{effective}
\end{equation}
%{effective}

\begin{figure}[htb]
\begin{center}
\unitlength=1.00mm
\special{em:linewidth 0.4pt}
\linethickness{0.4pt}
\begin{picture}(150.00,50.00)
\emline{10.00}{40.00}{1}{50.00}{40.00}{2}
\bezier{132}(13.00,10.00)(16.00,34.00)(22.00,40.00)
\bezier{32}(22.00,40.00)(25.00,43.00)(29.00,44.00)
\bezier{44}(29.00,44.00)(34.00,44.00)(40.00,42.00)
\bezier{40}(40.00,42.00)(46.00,41.00)(50.00,41.00)
\emline{10.00}{30.00}{3}{17.00}{30.00}{4}
\emline{60.00}{10.00}{5}{60.00}{50.00}{6}
\emline{60.00}{40.00}{7}{100.00}{40.00}{8}
\bezier{44}(63.00,50.00)(65.00,42.00)(67.00,40.00)
\bezier{52}(67.00,40.00)(71.00,33.00)(75.00,31.00)
\bezier{56}(75.00,31.00)(80.00,30.00)(89.00,33.00)
\bezier{44}(89.00,33.00)(91.00,34.00)(100.00,36.00)
\emline{73.00}{32.00}{9}{85.00}{32.00}{10}
\emline{71.00}{35.00}{11}{96.00}{35.00}{12}
\emline{68.00}{38.00}{13}{100.00}{38.00}{14}
\emline{68.00}{39.00}{15}{100.00}{39.00}{16}
\emline{110.00}{10.00}{17}{110.00}{50.00}{18}
\emline{110.00}{40.00}{19}{150.00}{40.00}{20}
\bezier{104}(110.00,15.00)(123.00,28.00)(130.00,30.00)
\bezier{84}(130.00,30.00)(137.00,33.00)(150.00,35.00)
\emline{110.00}{32.00}{21}{135.00}{32.00}{22}
\emline{110.00}{37.00}{23}{150.00}{37.00}{24}
\emline{110.00}{39.00}{25}{150.00}{39.00}{26}
\put(30.00,5.00){\makebox(0,0)[cc]{$a$}}
\put(80.00,5.00){\makebox(0,0)[cc]{$b$}}
\put(130.00,5.00){\makebox(0,0)[cc]{$c$}}
\emline{10.00}{39.00}{27}{21.00}{39.00}{28}
\emline{10.00}{37.00}{29}{20.00}{37.00}{30}
\emline{10.00}{10.00}{31}{10.00}{50.00}{32}
\emline{10.00}{15.00}{33}{14.00}{15.00}{34}
\emline{110.00}{20.00}{35}{115.00}{20.00}{36}
\put(22.00,25.00){\makebox(0,0)[cc]{${\cal U}(r)$}}
\put(78.00,25.00){\makebox(0,0)[cc]{${\cal U}(r)$}}
\put(132.00,25.00){\makebox(0,0)[cc]{${\cal U}(r)$}}
\put(31.00,47.00){\makebox(0,0)[cc]{${\cal U}_{\hskip0.5mm 0}$}}
\end{picture}
\caption{Effective potential and spectrum}
\label
{Eff}
\end{center}
\end{figure}
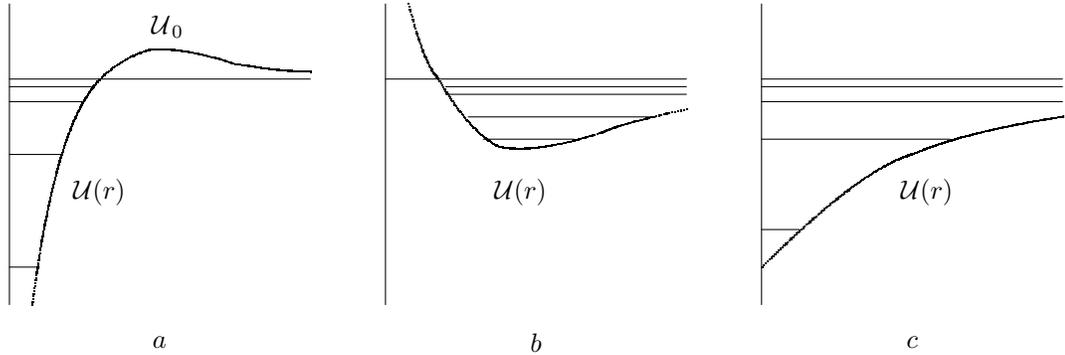

There are three alternatives. First, the attractive potential $U(r)$ is more
singular than the centrifugal term $J/r$. Figure \ref{Eff}$a$ depicts the
effective potential ${\cal U}(r)$. The particle could, in principle, orbit
in a circle of the radius corresponding to ${\cal U}_{\hskip0.5mm0}$, the
local maximum of ${\cal U}(r)$. But this orbiting is unstable, and the fall
to the centre is highly probable, not to mention the case $E>{\cal U}_{\hskip%
0.5mm0}$ when the fall to the centre is unavoidable.

Second, $U(r)$ is less singular than $J/r$. Specifically, for the potential $%
U(r)=-\alpha/r$ this means that $\alpha<J$. The shape of the ${\cal U}(r)$
curve is shown in Figure \ref{Eff}$b$. The particle executes a stable finite
motion. The fall to the centre is impossible, except when $J=0$ which is
associated with head-on collisions. But such collisions of point particles
have zero probability measure if dimensionality of the collision arena is
greater than 1.

Third, the singularities of $U(r)$ and $J/r$ are identical, i. e., $\alpha=J$. 
The effective potential ${\cal U}(r)$ is exhibited in Figure \ref{Eff}$c$.
The particle travels in a stable orbit that passes through the origin, but
this does not arrest the movement.

The quantum-mechanical treatment \cite{case} essentially confirms these
conclusions. It follows from the solutions of relativistic wave equations
for spin 0 and 1/2 particles that, in the case of sufficiently singular
potentials $U(r)$, bound states form a discrete spectrum\footnote{%
Surprisingly, the merge of two particles into a single one does not result
in a continuous spectrum signifying that the formed particle has no internal
structure and free. In quantum mechanics, the individuality of objects does
not preserve, and hence a superposition of states of two separate particles
and the aggregate of two merged particles is possible. A particle itself is
not fundamental, rather its degrees of freedom, in particular the energy
spectrum, and environmental features, e. g., the potential $U(r)$ are
fundamental concepts there.} extending from $E=m$ to $E=-\infty$; at $E=m$
there is a point of accumulation (Figure \ref{Eff}$a$). The system will tend
to more and more advantageous states associated with successively lower
energy levels. As this take place, the dispersion of the wave function
diminishes to zero when $E_n\to-\infty$. The process closely resembles the
fall to the centre in its classical interpretation. If $U(r)$ is less
singular than the properly defined quantum-mechanical centrifugal term, the
situation is conventional. The spectrum is bounded from below (Figure \ref
{Eff}$b$). The only distinctive feature of the quantum-mechanical situation
is that there exists a stable ground state with $J=0$. However, this does
not amount to the fall to the centre since the wave function behaves
smoothly in the vicinity of the origin; there is balance between attraction
and zero-point motion.

If singularities of $U(r)$ and the centrifugal term are identical, the
approach based on the one-particle wave equations becomes invalid. At short
distances, processes of the pair creation and annihilation enter the scene,
and methods of QFT are called for. As the coupling $\alpha$ tends to its
critical value $\alpha_c$, a highly nontrivial picture can arise which
continues to capture the theorists' attention (see, e. g., \cite{Kogut} and
references therein).

As to the qualitative results of this analysis, they are quite reliable to
suggest a criterion that separates systems where the fall to the centre is
suppressible from those where it is inevitable. The criterion reads: The
fall to the centre is preventable if and only if the energy spectrum is
bounded from below.

\section{Suppressibility of collapse}
\label{Suppress} For systems with infinite number of degrees of freedom, the
criterion should be extended to include them: {\it The tendency to collapse
is suppressible if and only if the spectrum is limited from below}. Indeed,
a negative contribution to the Hamiltonian is associated with attraction, and,
on the other hand, attractive forces result in the collapse if the energy 
spectrum extends to minus infinity.

However, the spectrum of much of studied systems can hardly be conceived
beforehand. The nonlinearity of dynamics can cause rearrangements of vacuum,
whence it follows that  na{i}ve realizations of the spectrum suggested by
the formal structure of Hamiltonian is far from true.

We thus have to change {\it the statement of the problem}. We can
confidently work with Hamiltonians quadratic in fields which in every case
of physical interest possess spectra bounded from below. We split the
Hamiltonian of the model at hand $H=\int \!d^{D-1}\!x\,{\cal H}(x)$ into two
parts $H=H_{0}+H_{I}$ where $H_{0}$ contains terms no greater than quadratic
in fields, $H_{I}$ containing the rest. $H_{0}$ is usually interpreted as
the quantity governing the free field evolution, $H_{I}$ being that
responsible for the interaction. As is well known, in the free field case,
the vacuum and one-particle states are stable, that is, vacuum fluctuations
have no impact on the spectrum. If one requires that the vacuum
root-mean-square fluctuation of $H_{0}$ be larger than that of $H_{I}$, then
the spectrum structure, in particular the boundedness from below, will be
left intact even in the presence of the interaction. Thus the tendency to
collapse is suppressed in such systems in which 
\begin{equation}
\Delta H_{0}>\Delta H_{I}.  \label{Delta}
\end{equation}
%(Delta)

The analogy with the Kepler problem is fairly close: The reason for keeping
the particle from the fall to the centre is that kinetic energy
manifestations (centrifugal effects or zero-point motion) dominate over
attractive forces.

Let us verify that (\ref{Delta}) is equivalent to the power-counting
criterion of the renormaliz\-abi\-lity. Because field operators in $H$ are
assumed to be normally ordered, the vacuum expectations $\langle
0|H_{0}|0\rangle $ and $\langle 0|H_{I}|0\rangle $ are vanishing, and (\ref
{Delta}) acquires the form 
\begin{equation}
\langle 0|\,H_{0}^{\hskip0.5mm2}|0\rangle >\langle 0|\,H_{I}^{\hskip%
0.5mm2}|0\rangle .  \label{H^2}
\end{equation}
%(H^2)
(Note, this inequality may be regarded as a necessary condition for
expanding in a perturbation series: For terms of the perturbation series to
decrease from order to order, the `unperturbed Hamiltonian' $H_{0}$ should
exceed the `perturbation' $H_{I}$ in some sense.)

Both sides of (\ref{H^2}) imply spatial integrations which generally result
in ultraviolet diver\-gences. The situation can be remedied in the usual
fashion by introducing a point-splitting. Furthermore, on account of the
positive definiteness of the integrands, it would be valid to compare not
the results of the integrations but only singularities of the matrix
elements $\langle 0\vert\,{\cal H}_0(x)\,{\cal H}_0(y)\vert 0\rangle$ and $%
\langle 0\vert\,{\cal H}_I(x)\,{\cal H}_I(y)\vert 0\rangle$ as $x^{\hskip%
0.1mm\mu}\to y^{\hskip0.1mm\mu}$.

The divergences emerging in this limit is evidence that the field quanta
behave singularly on the light cone. But this is not quite the issue of our
prime interest; a key role in the problem of collapse play short Euclidean
distances rather than short pseudo-Euclidean intervals. The ultraviolet
divergences originate from short Euclidean distances, if the time variable $%
x^{0}$ is changed to $ix^{0}$. To justify this change one employs the
analytic continuation which is possible provided that the matrix elements
are regularized in such a way that the rotation of the time axis $%
x^{0}\rightarrow ix^{0}$ does not intersect their singularities. Such is the
case, if the regularization includes not only the point-splitting but also a
chronological ordering of operator-valued fields.

The chronologically ordered multiplication is not the unique operation.
There are actually two alternative definitions of the $T$-product, by Dyson
and by Wick. One can adopt the Wick $T_{w}$-product (which is more
convenient being commutative with differentiations), accompanying this
choice by the replacement of Hamiltonian with Lagrangian \cite
{bogshirk,Pavlov}. Then, instead of (\ref{H^2}), we get 
\begin{equation}
\langle 0|\,T_{w}\{{\cal L}_{0}(x)\,{\cal L}_{0}(y)\}|0\rangle >\langle
0|\,T_{w}\{{\cal L}_{I}(x)\,{\cal L}_{I}(y)\}|0\rangle   \label{crit-reg}
\end{equation}
%(crit-reg)
where ${\cal L}_{0}$ and ${\cal L}_{I}$ are the Lagrangian densities
related, respectively, to ${\cal H}_{0}$ and ${\cal H}_{I}$. Dealing with
the situation in which free-field fluctuations dominate over those of
interaction, it is allowable to use the interaction picture where ${\cal L}%
_{0}$ and ${\cal L}_{I}$ depend of free fields.

Let the system be located in a flat $D$-dimensional spacetime and specified
by a set of $N$ real-valued fields generally symbolized as $\chi
_{j}(x),\,j=1,\ldots ,N$. (For now, let us ignore subtleties related to
constrained systems \cite{gtue}. They can reveal themselves in the final
result only implicitly through presence or lack of the gauge invariance.) We
set 
\begin{equation}
{\cal L}_{0}=\sum_{j=1}^{N}{\bf :}\chi _{j}(x)\,\L _{j}(\partial )\,\chi
_{j}(x){\bf :}  \label{example}
\end{equation}
%{example}
where $\L _{j}(\partial )$ is a differential operator of the first order for
fermions and of the second order for bosons, the symbol ${\bf :}\,{\bf :}$
stands for the normal product. Let us consider the simplest nontrivial
Lagrangian of interaction in the form  
\begin{equation}
{\cal L}_{I}=g\,\prod_{i=1}^{n}{\bf :}P^{k_{i}}(\partial )\,\chi _{i}(x){\bf %
:}\,  \label{L}
\end{equation}
%{L-I}
where $P^{k_{i}}(\partial )$ is a differential operator of $k_{i}$th order.

Substitution of (\ref{example}) and (\ref{L}) into (\ref{crit-reg}) yields
expressions built out of Feynman propagators $\Delta _{Fj}(x)$ analytically
continued to the $D$-dimensional Euclidean space. $\Delta _{Fj}(x)$
satisfies the equation 
\[
\L _{j}(\partial )\,\Delta _{Fj}(x)=-\delta ^{D}(x),
\]
and can be recast in the form 
\[
\Delta _{Fj}(x)=Q^{r_{j}}(\partial )\,\Delta _{F}(x).
\]
Here, $Q^{r_{j}}(\partial )$ is a differential operator of $r_{j}$th order
specific to the field of the given spin, $\Delta _{F}(x)$ is a kernel of the
operator $(\Delta +m_{j}^{2})^{-1}$, and $\Delta $ is the $D$-dimensional
Laplacian. Of concern to us is the situation at short Euclidean distances $%
x^{2}=\epsilon ^{2}$ where $\Delta _{F}(x)\sim x^{2-{D}}$, and 
\[
\Delta _{Fj}(x)\sim Q^{r_{j}}(\partial )\,x^{2-{D}}=O({\epsilon }^{\hskip%
0.5mm2-D-r_{j}}).
\]
For estimating the formal quantity $\delta ^{D}(0)$ one should introduce the
ultraviolet cutoff $\Lambda \sim 1/\epsilon $ in the Fourier integral 
\[
\delta ^{D}(x)=\frac{1}{(2\pi )^{D}}\int d^{D}k\,e^{ikx},
\]
then one obtains 
\[
\delta ^{D}(0)=O(\epsilon ^{-D}).
\]

With these estimations, in the limit $(x-y)^{2}=\epsilon ^{\hskip%
0.5mm2}\rightarrow 0$, we have 
\begin{equation}
\langle 0|\,T_{w}\{{\cal L}_{0}(x)\,{\cal L}_{0}(y)\}|0\rangle
=\sum_{j=1}^{N}\,\lbrack \L _{j}(\partial )\,\Delta _{Fj}(x-y)\rbrack ^{2}=O(%
{\epsilon }^{-2D}),  \label{L-0}
\end{equation}
%(L-0)
\begin{equation}
\langle 0|\,T_{w}\{{\cal L}_{I}(x)\,{\cal L}_{I}(y)\}|0\rangle
=g^{2}\prod_{i=1}^{n}P^{k_{i}}(\partial _{x})\,P^{k_{i}}(\partial
_{y})\,\Delta _{Fi}(x-y)=O(\prod_{i=1}^{n}{\epsilon }^{\hskip%
0.5mm2-D-r_{i}-2k_{i}}).  \label{L-I}
\end{equation}
%(L-I)
Comparing powers of $\epsilon $ in (\ref{L-0}) and (\ref{L-I}), we find that
the condition (\ref{crit-reg}) is met when 
\begin{equation}
\Omega _{n}\,=\,D+\sum_{i=1}^{n}\,(1-{\scriptstyle\frac{1}{2}}\,D-{%
\scriptstyle\frac{1}{2}}\,r_{i}-k_{i})\geq 0.  \label{crit-count}
\end{equation}
%(crit-count)
One can readily see that the inequality (\ref{crit-count}) is nothing but
the power-counting criterion for the renormalizability. For example, taking
into account that $r_{i}=0$ for scalar fields and $r_{i}=1$ for Dirac
fields,  for $D=4$, one finds $\Omega _{n}=-\omega _{n}$, where $\omega _{n}$
is defined by Eq.(\ref{omega_n}). Note also that the propagator of massive
vector fields $(\eta _{\mu \nu }-\partial _{\mu }\partial _{\nu
}/m^{2})\,\Delta _{F}$ behaves as $x^{-{D}}$ for  short distances, as
compared with the propagator of massless gauge fields $(\eta _{\mu \nu
}-\partial _{\mu }\partial _{\nu }/\partial ^{2})\,\Delta _{F}$, that
behaves as $x^{2-{D}}$, whence it follows that the criterion of
suppressibility of collapse (\ref{crit-count}) responds to presence or lack
of the gauge invariance through the value of $r_{i}$ specific to the vector
fields in question.

If (\ref{crit-count}) is an exact equality, the coupling $g$ becomes
dimensionless. As this take place, the meeting of the condition (\ref
{crit-reg}) is feasible when the magnitude of $g$ is sufficiently small.

The presented analysis must be considered as heuristic arguments of the
prevent\-ability of collapse. The basis for more rigorous treatment might
be, for example, properties of exact solutions to the Bethe--Salpeter
equation that describes quantum-field bound states. Unfortunately, efforts
of attacking this challenging problem (see, e. g., \cite{Halpern}) met with
little success for lack of adequate mathematical technique.

\section{Underlying reason for renormalizability}
\label{Renor} According to Dyson, a theory is renormalizable if all the
ultraviolet divergences are absorbed by a redefinition of parameters in the
Lagrangian. To fulfil this condition, it is necessary that vacuum
fluctuations of the kinetic energy exceed those of the interaction. Thus the
renormalizability can be substituted with the equivalent but more intuitive
concept of the {\it kinetic dominance}\footnote{%
D. I. Blokhintsev \cite{Blokhintsev} was the first to suggest this concept 
in somewhat simplified and implicit form.}.

It is a long-standing belief ({\small see, e. g., \cite{BSh}}) that, in
renormalizable theories, the physics at large distances is insensitive to
the influence of short distances and may be effectively allowed for by a
finite number of parameters, whereas such influence in the nonrenormalizable
case is given by infinite number of parameters. Our analysis supports this
idea. To be specific we turn back to the Kepler problem. In the situation
where the fall to the centre is suppressed (Figure \ref{Eff}$b$), the run of
the ${\cal U}(r)$ curve in the vicinity of the origin approaches to that of
the $I/r$ curve, hence the probe of short distances is controlled by the
only parameter $I$. By contrast, in the situation where the fall to the
centre is inevitable (Figure \ref{Eff}$a$), the run of the ${\cal U}(r)$
curve becomes close to that of the $U(r)$ curve as $r\to 0$, and, for the
probe of short distances, account must be taken of infinite number of
coefficients of the Laurent series representing $U(r)$.

Thus the renormalizability guarantees the {\it self-sufficiency of laws
driving large-scale phenomena}. However, it is not to be supposed that all
the phenomena occurring at the short-wave range are trivial, or at least can
be unified with those in the long-wave range. Due to  the suppression of
collapse, the low-energy region under study is in effect isolated from the
unexplored high-energy region.

It is well to bear in mind that the genuine condition of the preventability
of collapse is not the kinetic dominance but the boundedness of the spectrum
from below. Is this substitution of one condition for the other troublesome?

The kinetic dominance is {\it necessary} for the collapse to be suppressed%
\footnote{%
Curiously enough, the role of the energodominance in the problem of 
gravitational singularity is exactly opposite: Breaking the energodominance 
prevents the singularity formation \cite{Zeldovich}.}, but {\it not 
sufficient}. Indeed, the Yukawa
term ${\cal L}_{I}=ig{\bar{\psi}}\gamma _{5}\psi \phi $ in itself
exemplifies a nonrenormalizable Lagrangian for $D=4$ even though the
condition (\ref{crit-count}) is met; the renormalizability cannot be reached
unless the scalar self-interaction ${\cal L}_{I}=-\lambda \phi ^{4}$ is
added. Another well-known example is a theory of the Yang--Mills--Higgs
type which is in general nonrenormalizable due to the presence of the axial
anomaly. The anomaly violates the gauge invariance making worse  the 
ultraviolet  behavior of  the Yang--Mills fields, similar to that of
massive vector fields. These examples demonstrate the existence of systems
in which ultraviolet divergences defy their absorption, and hence the
collapse is not suppressed  despite the fulfillment of the inequality (\ref
{Delta}).

Arguing unsophisticatedly, one might even question the necessity of the 
kinetic dominance. For one thing, the impact of the apparently positive
definite disturbance ${\cal H}_{I}=\lambda \phi ^{2n}$, $\lambda >0$, on the
spectrum is at first glance quite harmless: The undisturbed energy levels 
seem to remain positive for any $n>1$. However, an important fact 
is overlooked that the coupling $\lambda $ can be profoundly altered in 
response to the vacuum polarization, in particular it can be opposite in sign 
to the bare coupling. Such a strong vacuum polarization is missing from 
two-dimensional
superrenormalizable theories \cite{Glimm}, so that $n$ may be arbitrary for $%
D=2$. As to the case $D=4$, the restriction $n\leq 2$ implied by the
condition (\ref{crit-count}) proves relevant here. Another doubt can be put
by a statement known as the `equivalence theorem' by which two quantum field
theories related by a nonlinear field transformation 
\begin{equation}
\chi =\xi +\xi ^{2}F(\xi )  \label{chi-Phi}
\end{equation}
%(crit-count)
have the same $S$ matrix. It follows that the collapse is suppressed not
only for polynomial Lagrangians satisfying the criterion (\ref{crit-count}),
but also for any others obtained from them by the nonlinear transformations (%
\ref{chi-Phi}). However, using the technology of nilpotent operators, as it
is applied in gauge field theories, one can show that the effect on ${\cal L}
$ generated by the nonlinear part of this transformation is similar to a
`gauge fixing term' \cite{Blasi}. Thus, the condition (\ref{Delta}) is in fact 
necessary for the suppression of collapse.

On the other hand, it may appear that the scalar self-interaction 
$\phi ^{\hskip0.5mm3}$ delivers a counter-example of the 4D renormalizable 
theory with the Hamiltonian spectrum unlimited from below. 
Indeed, the term $\phi ^{\hskip0.5mm3}$ tends to $-\infty $ 
more rapidly than the term ${\scriptstyle\frac{1}{2}}\,m^{2}
\phi ^{\hskip0.5mm2}$ tends to $\infty $ as $\phi\rightarrow -\infty$. 
In this consideration, however, the essential fact is overlooked; the
leading term in the kinetic energy is not ${\scriptstyle\frac{1}{2}}\,m^{2}
\phi ^{\hskip0.5mm2}$ but ${\scriptstyle\frac{1}{2}}\,(\partial \phi
)^{2}$ and the singularity of the latter, $O({\epsilon }^{-4})$, is comparable 
with the singularity of $\phi ^{\hskip0.5mm4}$ clearly exceding the
singularity of $\phi ^{\hskip0.5mm3}$.

The kinetic dominance criterion pertains to systems disposed to collapse.
Ultraviolet divergences warn of such disposition. The tendency to collapse
is suppressed in some systems while it is accomplished without obstacles in 
the rest systems. One can judge which alternative is inherent in the given 
system not only from analytical but also from topological features of its
behavior. The topological picture where the fall to the centre is the case
is qualified as ${\widehat{{\it 1}}}$-, ${\widehat{{\it 3}}}$-, and ${%
\widehat{{\it 5}}}$-deficient. This observation in the conventional local QFT 
can immediately be extended to more general theories (for example, M theory) as
follows. Let conversions of some spatially extended objects to another ones of
lesser dimension (e. g., contracting a string to a point, squeezing a
membrane to a string, flatting a soliton to a plane wave, etc.) be possible
on the classical level. Such conversions are reminiscent of the special
variant of evolution ${\widehat{{\it 4}}}$ where two initial particles are
interpreted as two ends of contracting open string. If these conversions are
invertible then every special variant of evolution is available, otherwise
the picture is deficient: The very existence of some special variant is
incompatible with the existence of the reverse variant. We take this
deficiency to be {\it topological definition of the feasibility of collapse}. 
Simply stated, this deficiency is equivalent to an incurable failure of the 
time reversal. In the quantum mechanics language, this
deficiency implies that the Hilbert spaces ${\rm H}_{{\rm in}}$ and ${\rm H}%
_{{\rm out}}$ are different, that is, the unitarity is violated. As is shown
in \cite{Aizenman}, the scalar theory with the Lagrangian of interaction $%
\phi ^{4}$ is trivial for $D>4$. This hints that nonrenormalizable theories
are unitary only in that the collapse is referred to
`pre-historic' times, and the relationship ${\rm H}_{{\rm in}}={\rm H}_{{\rm %
out}}$ is assured for such part of the system which survive the collapse 
occuring in the remote past and in the following does not experience 
interaction.

The suppressibility of collapse is a remedy regaining the reversibility. But
the cost for this is further depletion of the topological picture: Every
special variant disappears. The remedy may be such vigorous that no
realistic renormalizable theory with nontrivial $S$ matrix is available.
Although examples of nontrivial theories satisfying all the Wightman axioms
in 2D and 3D  spacetimes was shown to exist \cite{Glimm}, the possibility
that, say, the scalar self-interaction $\phi ^{4}$ in four-dimensional
spacetime, an important part of the Standard Model, corresponds to $S=1$
must not be excluded. This brings up the question: Is this depleted
topological picture an immanence of each renormalizable theory?

\section{Holographic principle and anomalies}
\label{Hologr} It is likely that the reader's attention was drawn on the fact 
that the {\it spacetime dimension} $D$ repeatedly come up in our discussion. 
For example, $D$ enters the criterion of suppressibi\-lity of collapse 
(\ref{crit-count}). Further still $D=2$ and $D=3$ flashed in relation to the
marvellous realm where the vacuum polarization is weak and the description 
is superrenormalizable. The very notion of the collapse proposed in the
preceding section is based on conversions of extended objects to objects
of lesser dimension. ({\small It is intringuing that, in supersymmetric 
theories where divergences are partially or completely cancelled, the 
ultraviolet happiness is due to effective reductions of  $D$. Indeed, as 
was shown by G. Parisi and N. Sourlas \cite{Parisi}, a supersymmetric 
theory on the graded manifold with ordinary coordinates $x_{1},\ldots,
x_{D}$ and the Grassmannian coordinates $\theta $ and $\bar{\theta}$ 
is equivalent to a nonsupersymmetric theory on the manifold with the
coordinates $x_{1},\ldots ,x_{D-2}$. In other words, the explicit 
realization of the supersymmetry calls for coordinates of {\it negative} 
dimensions such as $\theta $ and $\bar{\theta}$. Thus the supersymmetry is an 
elevator transporting us into lower dimensions.})

Another important to our theme notion is the {\it reversibility}. Up till
now, these two notions were disconnected. We would gain considerable insight 
into the subject, if we would received at our disposal some device enabling a 
linkage of physical pictures for different $D$. Such a device actually exists. 
This is the so-called holographic principle. It allows to regard the 
irreversibility as an anomaly that spoils the dualism between {\it quantum} 
and {\it classical} descriptions of realms of adjacent dimensions.

The holographic principle was first proclaimed by 't Hooft \cite{Hooft3} and
L. Susskind \cite{Susskind} in the context of black holes. According to this
principle, information on degrees of freedom inside a volume can be
projected onto a surface (also called screen) which encloses this volume.
For lack of the generally accepted formulation, we will turn to provisional
versions of this principle discussed in the literature. Their essence is as
follows: A classical theory (which includes gravity), describing phenomena
within a volume, can be formulated as a quantum theory of these phenomena
(without gravity) projected onto the boundary of this volume. In such a
form, the holographic principle was confirmed in Ref.\cite{Maldacena} where 
the consistency between semiclassical supergravity in an
anti-de Sitter space and quantum superconformal Yang--Mills theory on the
boundary of this space was revealed.

There is reason to believe that the holographic principle is valid also when
gravity is excluded. Then it can be simply realized through a remarkable 
dualism due to V. de Alfaro, S. Fubini and G. Furlan \cite{Alfaro}. They 
argued that the generating functional for Green's functions of Euclidean QFT 
in $D$ dimensions coincides with the Gibbs average for classical statistical
mechanics in $D+1$ dimensions. In other words, there exists a correspondence
(the AFF dualism) between the classical picture in a spacetime ${M}_{D+1}$
and the quantum picture in $D$-dimensional sections of ${\ M}_{D+1}$ at any
instants. Such secti\-ons play the role of $D$-dimensional screens
carrying the hologram of what happens in ${M}_{D+1}$.

We recall the idea of AFF dualism by the example of a system described by
the scalar field $\phi (x)$. Let the system be located in a $D$-dimensional
Euclidean spacetime and specified by the Lagrangian ${\cal L}$. One
introduces a fictitious time $t$. The field becomes a function of the
Euclidean coordinates $x_{1},\ldots ,x_{D}$ and fictitious time $t$, $\phi
=\phi (x,t)$. If ${\scriptstyle\frac{1}{2}}(\partial \phi /\partial t)^{2}$
is treated as the kinetic term, and ${\cal L}$ the potential energy term,
then one defines a new Lagrangian 
\[
{\tilde{{\cal L}}}={\scriptstyle\frac{1}{2}}(\partial \phi /\partial t)^{2}-%
{\cal L} 
\]
generating the evolution in $t$. The associated Hamiltonian is 
\[
{\tilde{{\cal H}}}={\scriptstyle\frac{1}{2}}\,\pi ^{2}+{\cal L} 
\]
where $\pi =\partial {\tilde{{\cal L}}}/\partial {\dot{\phi}}=\partial \phi
/\partial t$ stands for the conjugate momentum which is assumed to obey the
classical Poisson bracket 
\begin{equation}
\{\phi (x,t),\pi (y,t)\}=\delta ^{D}(x-y).  \label{P-B}
\end{equation}
% (P-B)
It is easy to see that the Gibbs average for an ensemble with the
temperature $kT=\hbar $ 
\begin{equation}
{\cal Z}\lbrack J\rbrack =\int {\cal D}\pi {\cal D}\phi \,\exp \Bigl(-\frac{1%
}{kT}\int d^{D}x\,({\tilde{{\cal H}}}+J\phi )\Bigr)  \label{Z-cl}
\end{equation}
% (Z-cl)
turns to the generating functional for the quantum Green functions 
\begin{equation}
Z\lbrack J\rbrack =\int {\cal D}\phi \,\exp \Bigl(-\frac{1}{\hbar }\int
d^{D}x\,({\cal L}+J\phi )\Bigr)  \label{Z-q}
\end{equation}
% (Z-q)
upon taking the Gaussian integral over $\pi $. Note that the holographic
mapping of the bulk picture in ${M}_{D+1}$ onto the screen picture in a
section of ${M}_{D+1}$ at any instant $t$ is ensured by the Liouville
theorem. Indeed, although $\phi (x,t)$ and $\pi (x,t)$ evolve in $t$, the
elementary volume in phase space ${\cal D}\pi {\cal D}\phi $, and with it
the Gibbs average ${\cal Z}$ are $t$-independent. ({\small For the AFF dual
description of gauge systems see \cite{Alfaro}.})

Thus, it is meaningless to ask whether a given realm is classical or
quantum. It may appear both as classical and quantum, but these two looks
pertain to spacetimes of nearby dimensions\footnote{%
By comparison, 't Hooft in his recent work \cite{Hooft99} claimed: ``In our
theory, quantum states are not the primary degrees of freedom. The primary
degrees of freedom are deterministic states.''}. To identify the realm, one
should only indicate $D$. For example, assigning $D=4$ to the
electromagnetic realm, we bear in mind that it can be grasped by either some
`quantum' 4D Lagrangian ${\cal L}$ or the associated `classical' 5D
Lagrangian ${\tilde{{\cal L}}}$.

It is clear that the conventional procedure of quantization only shifts the
seat to another realm: In lieu of the initial classical system living in $D$
dimensions, a new classical system living in $D+1$ dimensions emerges. As is
well known, symmetries of classical Lagrangians may be sensitive to the
dimension; some of them are feasible only for a single $D^\star$, while
another only for $D=2n$. On the other hand, given a quantized $D^\star$%
-dimensional theory, we deal actually with the holographic image of
classical $D^\star+1$-dimensional theory, and the symmetry under examination
is sure to be missing from it. Therein lies the reason for the symmetry
damage due to quantization known as the `quantum anomaly'. The
responsibility for this damage rests with the fact that the compared
classical descriptions merely differ in dimensions.

In the general case, the AFF dualism does not imply that the classical $D+1$%
-dimensional theory is {\it equivalent} to its quantum $D$-dimensional
counterpart. There are two reasons for the lack of equivalence. First, the
field theory can suffer from ultraviolet divergences, and the equality $%
{\cal Z}=Z$, strictly speaking, does not have mathematical sense. The
dissimilarity of classical divergences from quantum ones keeps the
expressions ${\cal Z}$ and $Z$ from being equal. This lack of equivalence
prohibits derivation of exact expressions for anomalies by a direct
comparison of two classical actions of adjacent dimensions. Furthermore, it
gives rise to a {\it classical anomaly} related to the violation of
reversibility on the classical level with preserving this symmetry on the
quantum level.

Second, given a one-to-one holographic mapping of a classical picture onto a
quantum one, one finds that the image displays a violation of the classical
determinism\footnote{%
Mathematically, the determinism is embodied in the requirement that the
solution to any Cauchy problem for classical dynamical equations must be
unique.} implying the loss of information on the classical system behavior.

Let us briefly run through these issues following essentially Ref. \cite{k0}.

\subsection{Conformal anomaly}

\label{anomalies} The holography is particularly suited for studying the
origin of anomalies since one may repeatedly handle classical actions alone.
Canonical transformations leave the Poisson bracket (\ref{P-B}) invariant,
hence the measure of integration in (\ref{Z-cl}) ${\cal D}\pi{\cal D}\phi$
has a corresponding canonical invariance property enabling one to perform
canonical transformations without having to introduce a Jacobian. This can
be exemplified by the conformal anomaly in the Yang--Mills theory.

Let us consider the conformal transformation of the metric 
\begin{equation}
g_{\mu\nu}\to e^{2\varepsilon}\,g_{\mu\nu}  \label{conf-transf}
\end{equation}
% (conf-transf)
resulting in the associated Noether current, the energy-momentum tensor 
\begin{equation}
\Theta^{\mu\nu}=\frac{2}{\sqrt{- g}}\,\frac{\delta}{\delta g_{\mu\nu}} \,%
\sqrt{- g}\,{\cal L}.  \label{Theta-def}
\end{equation}
% (Theta-def)
The conformal invariance of the classical action $S$ is attained if $\delta
S=2\varepsilon\,g_{\mu\nu}\Theta^{\mu\nu}=0$, that is, 
\begin{equation}
\Theta^{\mu}_{\hskip1mm\mu}= 0.  \label{Theta-mu-mu}
\end{equation}
% (Theta-mu-mu)
It follows from (\ref{Theta-def}) that, in the classical $D+1$-dimensional
Yang--Mills theory with 
\[
{\cal L}_{{\rm YM}}=-\frac{1}{4\Omega_{D-1}\alpha}\,{\rm tr}%
\,F_{\alpha\beta} F^{\alpha\beta}, 
\]
$\Theta^{\mu\nu}$ is written as 
\begin{equation}
\Theta^{\mu\nu}=\frac{1}{\Omega_{D-1}\alpha}\,{\rm tr}\,(F^{\mu\alpha}
F_{\alpha}^{\hskip1.5mm\nu}+ {\frac{1}{4}}\,\eta^{\mu\nu}F_{\alpha\beta}F^{%
\alpha\beta}).  \label{Theta}
\end{equation}
% (Theta)
Here, $\Omega_{D-1}$ is the area of a $D-1$-dimensional unite sphere, and $%
\alpha$ the Yang--Mills coupling. The expression (\ref{Theta}) makes it
clear that the condition (\ref{Theta-mu-mu}) is fulfilled only for $D+1=4$.
Yang--Mills equations are conformally invariant only in a four-dimensional
spacetime. This is in agreement with the scale invariance of the theory
manifested in the fact that the coupling $\alpha$ is dimensionless for $%
D+1=4 $.

With the holographic principle, the quantization of the classical 4D
Yang--Mills theory culminates in the classical 5D Yang--Mills theory where
one finds $\Theta^{\mu}_{\hskip1mm\mu}\ne 0$. This is how the conformal
anomaly occurs.

Technically, the conformal invariance breakdown in QFT is traced to the mass
scale $\mu$ introduced for the normalization of $\alpha$. S. Coleman and E.
Weinberg \cite{Coleman} called this the dimensional transmutation. Such a
transmutation is missing from the quantum 3D Yang--Mills theory, thereby the
conformal invariance of its AFF dual, the classical 4D Yang--Mills theory,
is ensured. Why does this happen? In the next section, we will see that the
quantum 3D theory is super-renormalizable, the vacuum polarization is weak,
there is no infinite charge renormalization, and the need for $\mu$
disappears.

As to the quantum 4D theory, vacuum polarization effects are significant
here, and $\alpha$ becomes the running coupling constant $\alpha(q^2/\mu^2)$
dependent on the momentum transfer squared $q^2$, implying that 
\begin{equation}
\Theta^{\mu}_{\hskip1mm\mu}=\frac{2g^{\mu\nu}}{\alpha^2}\,\frac{\partial
\alpha}{\partial g^{\mu\nu}}\,\frac{1}{16\pi}\,{\rm tr}\,F^2= \frac{1}{8\pi}%
\,\frac{\beta(\alpha)}{\alpha^2}\,{\rm tr}\,F^2  \label{Theta-mu-mu-q}
\end{equation}
% (Theta-mu-mu-q)
where $\beta=\partial\alpha/\partial\log q^2$ is the Gell-Mann -- Low
function. We encounter the $q^2$-dependent coefficient of ${\rm tr}\,F^2$ as
opposed to the na{\H\i}ve expression 
\begin{equation}
\Theta^{\mu}_{\hskip1mm\mu}=\frac{1}{8\pi^2\alpha}\,{\rm tr}\,F^2
\label{Theta-mu-mu-cl}
\end{equation}
% (Theta-mu-mu-cl)
which might be derived from the classical 5D Yang--Mills action. Thus the
holographic principle qualitatively explains the relation $\Theta^{\mu}_{%
\hskip1mm\mu}\propto {\rm tr}\,F^2$, but the exact coefficient of the
expression (\ref{Theta-mu-mu-q}) does not reproduce.

\subsection{Irreversibility}

\label{irreversib} The violation of the AFF equivalence after the infinite
renormalization can be regarded as a kind of anomalies. As an example, let
us consider the emergence of the irreversibility in a classical picture,
with the reversibility in the dual quantum picture being left intact. To be
specific, take a $D$-dimensional quantum realm described by the scalar
electrodynamics Lagrangian 
\begin{equation}
{\cal L}=(\partial^\mu+ig_0 A^\mu)\phi\,(\partial_\mu-ig_0 A_\mu){\bar {\phi}%
}-m_0^2\phi\,{\bar{\phi}}-\frac{\lambda_0}{4}(\phi\,{\bar{\phi}})^2 -\frac{1%
}{4\Omega_{D-2}}\,F_{\mu\nu}F^{\mu\nu}.  \label{L-scal}
\end{equation}
% (L-scal)

In the dual $D+1$-dimensional classical realm, the scalar field $\phi$ can
be treated as a Lagrangian coordinate of a continuous medium that evolves in
time $t=x_{D+1}$. However, such a model is inconvenient for analysis of
ultraviolet properties, and we will discuss its discrete analogue, the
system with very large number (strictly speaking, $\infty^D$) of charged
point particles. Assume that the action for the classical particle
interacting with electromagnetic field is of the usual form 
\begin{equation}
S=-\int d\tau\,(m_0\sqrt{v\cdot v}+ g\,v\cdot A)  \label{action}
\end{equation}
% (action)
where $v^\mu\equiv{\dot z}^\mu\equiv dz^\mu/d\tau$ is the $D+1$-velocity of
the particle, and $\tau$ the proper time.

Features of dual AFF pairs corresponding to several values of $D$ are
summarized in Table 1. Let us begin with the line `Ultraviolet behavior'
indicating the dependence of observables on the cutoff $\Lambda$. The
maximal power of $\Lambda$ increases with $D$ to give the progressive
violation of the AFF dualism up to its complete failure above $D=4$.

\begin{center}
\begin{tabular}{|c|c||c|c||c|c||c|c|}  
\multicolumn{8}{c}{TABLE 1.\quad{\bf  Dualities in scalar electrodynamics}}\\[2mm]\hline 
\multicolumn{8}{|c|}{ {\it Spacetime dimension of the quantum picture
}}\\ \hline
\multicolumn{2}{|c||}{$D=1$} & \multicolumn{2}{|c||}{$D=2$} & \multicolumn{2}{|c||}{$D=3$}&\multicolumn{2}{|c|}{$D=4$}   \\ \hline\hline
\multicolumn{8}{|c|}{%\small 
\it Dual AFF pairs}\\ \hline
\multicolumn{2}{|c||}{1D$_{\rm quant}$ -- 2D$_{\rm class}$} & \multicolumn{2}{|c||}{2D$_{\rm quant}$ -- 3D$_{\rm class}$} & \multicolumn{2}{|c||}{3D$_{\rm quant}$ -- 4D$_{\rm class}$}&\multicolumn{2}{|c|}{4D$_{\rm quant}$ -- 5D$_{\rm class}$}\\ \hline\hline
\multicolumn{8}{|c|}{%\small 
\it Ultraviolet behavior}\\ \hline
$\Lambda^0$ & $\Lambda^0$ & $\log\Lambda$ & $\log\Lambda $  & $\Lambda$ & $\Lambda$ & $\Lambda^2$, $\log\Lambda$ & $\Lambda^2$, $\log\Lambda$ \\ 
\hline\hline
\multicolumn{8}{|c|}{%\small 
\it Renormalizability}\\ \hline
{\small  finite} &{\small  finite}  & {\small super-}    & {\small ren'zable}  & {\small super-}    & {\small ren'zable} & {\small ren'zable} & {\small non-} \\ 
{\small        } &{\small        }  & {\small ren'zable} & {\small }         & {\small ren'zable} & {\small          } &{\small           } & {\small ren'zable}\\ \hline\hline
\multicolumn{8}{|c|}{%\small 
\it Reversibility}\\ 
\hline%\hline
{\small holds}     & {\small holds}     & {\small holds}     & {\small holds    } & {\small holds   }   & {\small weak}  & {\small holds}& {\small topolog}\\
                 &                  &                  &                  &                   & {\small violat}  &             &  {\small violat} \\\hline%\hline
\end{tabular}
\end{center}

Among $D+1$-dimensional classical quantities, the dependence on $\Lambda$ is
inherent in the energy-momentum vector\footnote{%
The angular momentum has the same ultraviolet behavior} of electromagnetic
field generated by a point particle, 
\begin{equation}
P_\mu=\int d\sigma^\nu\,\Theta_{\mu\nu},  \label{P-mu}
\end{equation}
% (P-mu)
where the integration is performed over a $D$-dimensional spacelike
hypersurface. In the static case when electromagnetic field can be specified
by the potential $\varphi$ satisfying the Poisson equation 
\begin{equation}
\Delta\varphi ({\bf x})=-\Omega_{D-1}\,\rho({\bf x})  \label{Poisson}
\end{equation}
% (Poisson)
with 
\begin{equation}
\rho({\bf x})=g\,\delta^{D}({\bf x}),  \label{rho}
\end{equation}
% (rho)
one has 
\begin{equation}
\varphi({\bf x})=g \cases{\vert\,{\bf x}\vert^{2-D}, & $D\ne 2$,\cr
\log\vert\,{\bf x}\vert, & $D=2$.\cr}  \label{varphi}
\end{equation}
% (varphi)
From (\ref{rho}) and (\ref{varphi}) in combination with the static
expression for the particle self-energy 
\begin{equation}
\delta m={\frac12}\int d^{D}{\bf x}\,\rho({\bf x})\,\varphi({\bf x})\, ={%
\frac12}\,g^2\varphi(0),  \label{delta m}
\end{equation}
% (delta m)
one obtains leading $\Lambda$-dependences indicated in Table 1.

For small deviations from statics, Eq. (\ref{P-mu}) acquires the form 
\begin{equation}
P_\mu=c_1\,v_\mu+ c_2\,{\dot v}_\mu +c_3\,{\ddot v}_\mu +\ldots
\label{quasi}
\end{equation}
% (quasi)
where the variables ${v}_\mu$, ${\dot v}_\mu$, etc., relate to the point
source of the field at a given instant. It is clear that $c_1=\delta m$.
From dimensional considerations, one finds also that 
$c_i/c_{i+1}\sim\Lambda$.

Let us turn to 5D$_{{\rm class}}$. The integration of Eq. (\ref{P-mu}) is
carried out over a four-dimensional hypersurface, and, therefore, only even
powers of $\Lambda$ are nonzero: 
\begin{equation}
c_1\sim\Lambda^2,\quad c_2=0,\quad c_3\sim\log\Lambda.  \label{c-i}
\end{equation}
% (c-i)
The $\Lambda^2$ term is absorbed by the mass renormalization, but there is
no parameter in the action (\ref{action}) suitable for the absorption of the
logarithmic divergence, and the 5D$_{{\rm class}}$ theory turns out to be 
{\it nonrenormalizable}. To regain the renormalizability, to the action (\ref
{action}) must be added a term with higher derivatives \cite{k9} of the type 
\begin{equation}
-\kappa_0 \int d\tau\, \frac{1}{\sqrt{{v}\cdot{v}}}\,\Biggl(\frac{d}{d\tau}%
\frac{{v}^\mu} {\sqrt{{v}\cdot{v}}}\Biggr)^2.  \label{222}
\end{equation}
% (222)

\begin{figure}[htb]
\begin{center}
\unitlength=1.00mm
\special{em:linewidth 0.4pt}
\linethickness{0.4pt}
\begin{picture}(112.00,18.00)
\emline{10.00}{10.00}{1}{12.00}{10.00}{2}
\emline{13.00}{10.00}{3}{15.00}{10.00}{4}
\emline{25.00}{10.00}{5}{27.00}{10.00}{6}
\emline{28.00}{10.00}{7}{30.00}{10.00}{8}
\emline{36.00}{10.00}{9}{38.00}{10.00}{10}
\emline{39.00}{10.00}{11}{41.00}{10.00}{12}
\emline{42.00}{10.00}{13}{44.00}{10.00}{14}
\emline{52.00}{10.00}{15}{56.00}{10.00}{16}
\emline{66.00}{10.00}{17}{70.00}{10.00}{18}
\emline{76.00}{10.00}{19}{84.00}{10.00}{20}
\emline{92.00}{12.00}{21}{97.00}{10.00}{22}
\emline{97.00}{10.00}{23}{92.00}{8.00}{24}
\emline{107.00}{10.00}{25}{112.00}{12.00}{26}
\emline{112.00}{8.00}{27}{107.00}{10.00}{28}
\bezier{56}(15.00,10.00)(20.00,15.00)(25.00,10.00)
\bezier{56}(15.00,10.00)(20.00,5.00)(25.00,10.00)
\bezier{56}(40.00,10.00)(35.00,17.00)(40.00,18.00)
\bezier{56}(40.00,10.00)(45.00,17.00)(40.00,18.00)
\bezier{56}(56.00,10.00)(61.00,15.00)(66.00,10.00)
\bezier{16}(59.00,8.00)(61.00,7.00)(63.00,8.00)
\bezier{16}(78.00,16.00)(80.00,17.00)(82.00,16.00)
\bezier{12}(77.00,15.00)(77.00,14.00)(78.00,13.00)
\bezier{12}(83.00,15.00)(83.00,14.00)(82.00,13.00)
\bezier{16}(100.00,12.00)(102.00,13.00)(104.00,12.00)
\bezier{16}(100.00,8.00)(102.00,7.00)(104.00,8.00)
\put(15.00,10.00){\makebox(0,0)[cc]{$\bullet$}}
\put(25.00,10.00){\makebox(0,0)[cc]{$\bullet$}}
\put(40.00,10.00){\makebox(0,0)[cc]{$\bullet$}}
\put(33.00,10.00){\makebox(0,0)[cc]{$+$}}
\put(33.00,3.00){\makebox(0,0)[cc]{$\Pi_{\mu\nu}$}}
\put(56.00,10.00){\makebox(0,0)[cc]{$\bullet$}}
\put(66.00,10.00){\makebox(0,0)[cc]{$\bullet$}}
\put(73.00,10.00){\makebox(0,0)[cc]{$+$}}
\put(80.00,10.00){\makebox(0,0)[cc]{$\bullet$}}
\put(73.00,3.00){\makebox(0,0)[cc]{$\Sigma$}}
\put(97.00,10.00){\makebox(0,0)[cc]{$\bullet$}}
\put(107.00,10.00){\makebox(0,0)[cc]{$\bullet$}}
\put(102.00,3.00){\makebox(0,0)[cc]{$\Upsilon$}}
\bezier{8}(80.00,10.00)(80.00,11.00)(79.00,12.00)
\bezier{8}(80.00,10.00)(80.00,11.00)(81.00,12.00)
\bezier{8}(97.00,10.00)(98.00,10.00)(99.00,11.00)
\bezier{8}(97.00,10.00)(98.00,10.00)(99.00,9.00)
\bezier{8}(107.00,10.00)(106.00,10.00)(105.00,11.00)
\bezier{8}(107.00,10.00)(106.00,10.00)(105.00,9.00)
\bezier{8}(56.00,10.00)(57.00,10.00)(58.00,9.00)
\bezier{8}(66.00,10.00)(65.00,10.00)(64.00,9.00)
\end{picture}
\caption{One-loop diagrams in scalar QED}
\label
{diagram}
\end{center}
\end{figure}
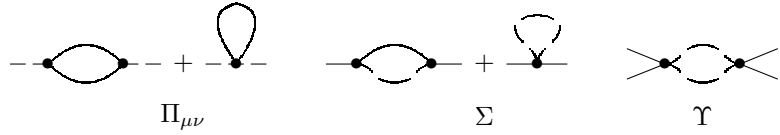

As to the dual $D$-dimensional quantum quantities, we are directly concerned
with the polarization operator $\Pi_{\mu\nu}$, the scalar self-energy $%
\Sigma $, and the scalar self-interaction $\Upsilon$. The relevant one-loop
diagrams are depicted in Figure \ref{diagram}. 
(The light-light scattering
revealing a more soft ultraviolet behavior is immaterial here.) Using a
gauge-invariant regularization, $\Pi_{\mu\nu}$ can be cast as $%
\Pi_{\mu\nu}(q)=(q^2\eta_{\mu\nu}-q_\mu q_\nu)\,\Pi(q^2)$. An elementary
Feynman technique leads to 
\[
\Sigma\sim\int\frac{d^Dk}{k^2}, \qquad \Pi\sim\int\frac{d^Dk}{(k^2)^2},
\qquad \Upsilon\sim\int\frac{d^Dk}{(k^2)^2}. 
\]
It follows 
\begin{equation}
\Sigma\sim \cases{\Lambda^{D-2}, & $D\ne 2$,\cr \log\Lambda, & $D=2$,\cr}
\qquad \Pi\sim \cases{\Lambda^{D-4}, & $D\ne 4$,\cr \log\Lambda, & $D=4$,\cr}
\qquad \Upsilon\sim \cases{\Lambda^{D-4}, & $D\ne 4$,\cr \log\Lambda, &
$D=4$.\cr}  \label{UV}
\end{equation}
% (UV)

The comparison of (\ref{varphi})--(\ref{delta m}) and (\ref{c-i}) with (\ref
{UV}) shows that the divergence powers in the 5D$_{{\rm class}}$ theory are
the same as those in the 4D$_{{\rm quant}}$ theory. Nevertheless, the latter
is {\it renormalizable} \cite{Salam} since all the primitively divergent
diagrams refer to the quantities $\Pi$, $\Sigma$ and $\Upsilon$ which
renormalize, respectively, $g_0$, $m_0$ and $\lambda_0$ in the Lagrangian (%
\ref{L-scal}).

Since the bar Lagrangian ${\cal L}_0$ and counter-terms in the 4D$_{{\rm %
quant}}$ theory have the same structure, and ${\cal L}_0$ is invariant under
time reversal $t\to -t$, the renormalized quantum dynamics is reversible.
Meanwhile the dual 5D$_{{\rm class}}$ theory is irreversible due to the
non-suppressed tendency to collapse (unless the action is modified {\it ad
hoc} by the addition of terms with higher derivatives). Indeed, in the
Kepler problem, the potential energy $U(r)=-g^2/r^{2}$ is more singular than
the centrifugal term $J/r$, and the fall to the centre is inevitable. The
fall to the centre is an irreversible phenomenon. Note, the irreversibility
has topological character: The mere presence of the variant of evolution ${%
\widehat{{\it 4}}}$ (the fall to the centre) rules out the possibility of
the inverse variant ${\widehat{{\it 3}}}$ (splitting and subsequent
departure of the merged particles). The violation of the AFF dualism 4D$_{%
{\rm quant}}$ -- 5D$_{{\rm clas}}$ is due to the fact that the classical
realm is devoid of the vacuum polarization which is responsible for infinite
renormalization of $g_0$ and $\lambda_0$.

Let us turn to the case $D=3$. Now $\Pi$ and $\Upsilon$ are finite, hence
the coupling constants are not subject to infinite renormalization. The
divergences of both $\Sigma$ and $\delta m$ are absorbed by the mass
renormalization. The number of parameters contained in the Lagrangians are
sufficient for the removal of all divergences, so, at first glance, the AFF
equivalence is the case. But this is wrong. The mass renormalization makes a
finite {\it mark} which is more deep in the classical picture than in the
quantum one. The renormalized quantum dynamics is reversible. By contrast,
in the renormalized classical dynamics, the reversibility is violated.
Indeed, the coefficients $c_i$ in (\ref{quasi}) are \cite{teit} 
\[
c_1\sim \Lambda,\quad c_2=-{\frac23}\,g^2,\quad c_3\sim \Lambda^{-1}, 
\]
whence it follows the expression for the four-momentum of the `dressed'
particle 
\begin{equation}
p_\mu=m\,v_\mu-{\frac23}\,g^2\,{\dot v}_\mu  \label{p-mu}
\end{equation}
% (p-mu)
($m$ is the renormalized mass), and the equation of motion of the `dressed'
particle 
\begin{equation}
m\,{\dot v}^\mu-{\frac23}\,g^2\,({\ddot v}^\mu+ {\dot v}^2 v^\mu)= f^\mu.
\label{LD}
\end{equation}
% (LD)
This is the Lorentz--Dirac equation which is evidently not invariant under
time reversal $\tau\to -\tau$. This irreversibility is due to the energy
dissipation stemming from radiation of electromagnetic waves.

{\small Many attempts to derive Eq. (\ref{LD}) from some Lagrangian have not
met with success. It is likely that the partition function (\ref{Z-cl})
whose construction is based on the Hamiltonian (which in turn implies the
availability of the corresponding Lagrangian) has nothing to do with the
renormalized classical dynamics. The irreversibility of the classical
evolution in the four-dimensional realm is attributable not to ergodic
properties but to the peculiarity of the self-interaction in classical
electrodynamics that the field energy can only be emitted (not absorbed)
that is, only dissipation (not cumulation) of energy is possible there. The
irreversibility is an anomaly revealing itself in the violation of the AFF
equivalence 3D$_{{\rm quant}}$ -- 4D$_{{\rm clas}}$.}

Thus a trait of the 4D$_{{\rm class}}$ theory is that the renormalization
deliver it from the invariance under time reversal which would contradict to
the macroscopic experience. It is significant that this excess invariance is
not violated at the expense of non-invariant terms of the Lagrangian, since
their presence would impair the invariance of the 3D$_{{\rm quant}}$ theory.
The weak violation of the AFF equivalence 3D$_{{\rm quant}}$ -- 4D$_{{\rm clas%
}}$ is adequate to the physical reality. It can be liken to the chiral
anomaly making possible the decay $\pi^0\to 2\gamma$ which is forbidden due
to the excess symmetry of the Standard Model.

In the case $D=2$, the ultraviolet situation is qualitatively the same as in
the case $D=3$: The divergences of both $\Sigma$ and $\delta m$ are absorbed
by the mass renormalization. However, the renormalized 3D$_{{\rm class}}$
dynamics is reversible. Indeed, $c_1\sim\log\Lambda$, therefore, $%
c_2\sim\Lambda^{-1}$. This means that the mass renormalization makes no
finite mark in the three-momentum of the `dressed' particle governed by the
second Newton law. Thus the AFF equivalence 2D$_{{\rm quant}}$ -- 3D$_{{\rm %
clas}}$ is observed.

In the case $D=1$, there are no ultraviolet divergences at all, and the AFF
equivalence cannot be violated for this reason. This case should be
discussed more elaborately.

\subsection{Indeterminism in a classical realm}

\label{special} Ultraviolet divergences is not the only reason why the AFF
dualism is not an exact equivalence. Another reason is that classical
pictures are devoid of quantum coherence, in other words, the classical
determinism cannot be reconciled with the quantum principle of
superposition. The notion of probability is incorporated in classical
statistical mechanics quite artificially; it expresses the measure of our
ignorance of deterministic pictures in detail, whereas the probability
amplitude is a fundamental element of quantum theory. The holographic
mapping of a classical theory onto a quantum theory should suffer
information loss, but the mechanism of this loss is obscure\footnote{%
To circumvent this difficulty, 't Hooft \cite{Hooft99} suggested that
``Since, at a local level, information in these (classical) states is not
preserved, the states combine into equivalence classes. By construction
then, the information that distinguishes the different equivalence classes
is absolutely preserved. Quantum states are equivalence classes.''}. We will
give a schematic reasoning demonstrating that this problem is tractable in
the case $D=1$, namely, a kind of indeterministic behavior of particles is
possible in the two-dimensional classical realm.

Let us write the dynamical equations of the discussed 2D$_{{\rm class}}$
theory: 
\begin{equation}
\partial_\lambda F^{\lambda\mu}(x)=2e \sum_{a=1}^2\int_{-\infty}^\infty
d\tau_a\, {v}^\mu_a(\tau_a)\,\delta^{(2)}\Bigl(x-z(\tau_a)\Bigr),
\label{maxw}
\end{equation}
% (maxw)
\begin{equation}
m_a {\dot v}^{\mu}_a=e_a {v}_{\nu}^a F^{\mu\nu}(z_a).  \label{newt}
\end{equation}
% (newt)
They are exactly integrable. From the form of solutions to these equations,
two striking features of the 2D$_{{\rm class}}$ realm follow. First, there
is no radiation of electromagnetic waves in this realm \cite{k9} and hence
there is no dissipation of energy. All motions of particles are reversible.
Second, it is possible that several point particles merge into a single
aggregate, and then, after a lapse of some period, split into the initial
objects.

For simplicity, we restrict our consideration to a system of two particles
with equal masses $m$ and charges $e$ (denoting $e^2/m=a$) which are to be
related to the centre-of-mass frame. Let the particles be moving towards
each other, and their total energy is such that, at the instant of their
meeting, their velocities are exactly zero. Then there exists an exact
solution to Eqs. (\ref{maxw}) and (\ref{newt}) describing two worldlines $%
z^\mu_1(\tau)$ and ${\ z}^\mu_2(\tau)$ which coalesce at the instant $%
\tau^\ast$ and separate at the instant $\tau^{\ast\ast}=\tau^{\ast}+T$, 
\begin{equation}
z^\mu_1(\tau)=\cases{a^{-1}\{{\sinh} a(\tau-\tau^\ast),1-{\cosh} a(\tau-
\tau^\ast)\},& $\tau<\tau^{\ast}$,\cr \{\tau-\tau^{\ast},0\}, &
$\tau^{\ast}\le\tau<\tau^{\ast\ast}$,\cr a^{-1}\{aT+{\sinh}
a(\tau-\tau^{\ast\ast}), {\cosh} a(\tau-\tau^{\ast\ast}) -1\},&
$\tau\ge\tau^{\ast\ast}$\cr},  \label{z}
\end{equation}
% (z^\mu)
\begin{equation}
z^\mu_2(\tau)=\cases{a^{-1}\{{\sinh} a(\tau-\tau^\ast),{\cosh} a(\tau-
\tau^\ast)-1\},& $\tau<\tau^{\ast}$,\cr z^\mu_1(\tau), &
$\tau^{\ast}\le\tau<\tau^{\ast\ast}$,\cr a^{-1}\{aT+{\sinh}
a(\tau-\tau^{\ast\ast}), 1-{\cosh} a(\tau-\tau^{\ast\ast})\},&
$\tau\ge\tau^{\ast\ast}$,\cr}  \label{-z}
\end{equation}
% (-z^\mu)
and the retarded field $F^{\mu\nu}$ expressed through $z_1^\mu(\tau)$ and ${%
\ z}_2^\mu(\tau)$ as 
\begin{equation}
F^{\mu\nu}(x)=e\,\sum_{a=1}^2\,\frac{1}{\rho_a}\,(R_a^{\mu} {v}_a^{\nu}-
R_a^{\nu} {v}_a^{\mu}).  \label{F}
\end{equation}
Here, $R^\mu_a\equiv x^{\mu}-z^{\mu}_{a\hskip0.3mm{\rm ret}}$ is an
isotropic vector drawn from the emitting point $z^{\mu}_{a\hskip0.3mm{\rm ret%
}}$ on $a$th worldline to the point of observation $x^\mu$, and $%
\rho_a=R_a\cdot {v}_a$ the invariant retarded distance between $x^\mu$ and $%
z^{\mu}_{a\hskip0.3mm{\rm ret}}$. %(F)   

The parameters $\tau^{\ast}$ and $\tau^{\ast\ast}$ are arbitrary. If $%
\tau^{\ast}$ and $\tau^{\ast\ast}$ are different and finite, then the curves
(\ref{z}) and (\ref{-z}) correspond to the formation of an aggregate with
finite life time, the variant ${\widehat{{\it 5}}}$. For $\tau^{\ast\ast}\to
\infty$, they describe the formation of a stable aggregate never decaying,
the variant ${\widehat {{\it 4}}}$. For $\tau^{\ast}\to -\infty$, we see
decay at a finite instant of an aggregate formed at the infinitely remote
past, the variant ${\widehat {{\it 3}}}$. If $\tau^{\ast}\to -\infty$ and $%
\tau^{\ast\ast}\to\infty$, then the curves degenerate into a straight line
corresponding to an absolutely stable aggregate, the variant ${\widehat{{\it %
2}}}$. For $\tau^{\ast}=\tau^{\ast\ast}$, they describe an aggregate which
exists for a single instant, the variant ${\widehat{{\it 1}}}$. Thus the
solution to Eqs. (\ref{maxw}) and (\ref{newt}) with the given Cauchy data is
not unique. Moreover, we have a continuum of solutions since the period of
the merged state can be any $T\ge 0$. The decay occurs quite accidentally at
any instant. Note, however, that variants of evolution with such Cauchy data
comprise null set.

In summary, a $D+1$-dimensional classical picture can be equivalent to the
AFF associated $D$-dimensional quantum picture if and only if $D=1$. Indeed,
the 1D$_{{\rm quant}}$ theory is free of ultraviolet divergences. It is just
the two-dimensional classical realm in which radiation is absent, the problem 
of dissipation of energy does not arise, and hence there is no anomaly of 
reversibility. Only in this realm, the retarded electromagnetic field is not 
singular on world lines of its sources, Eq.(\ref{F}), and the special variant
of evolution ${\widehat{{\it 4}}}$ must not be identified with the fall to
the centre. Besides, only this realm is compatible with the special variants 
of evolution rendering dynamics stochastic already on the classical level (in 
higher dimensions, special variants of evolution either comprise a
deficient picture or disappear altogether). Since the measure 
of such variants in phase space is zero, they
have no effect on the fulfillment of the Liouville theorem, so that the
variable $t$ is outside the quantum description where survives only $ix_{0}$
playing the role of the `Euclidean' time. Taking into account that clusters
of classical charged particles in the two-dimensional realm mimic classical
strings, it becomes clear that the behavior of classical strings is coded in
the behavior of quantum point objects.

\section{Conclusion}
\label{Conclusion} We have endeavored to argue that the renormalizability
amounts to the kinetic domi\-nan\-ce. The latter is necessary for
suppressing the tendency to collapse. Why do we worry about the collapse?
The picture where the collapse occurs is irreversible, with the
irreversibility being of topological origin. Such a picture is
topolo\-gically deficient; in the quantum mechanical language, the spaces of
asymptotic states ${\rm H}_{{\rm in}}$ and ${\rm H}_{{\rm out}}$ differ, so
that the unitarity is flawed at least on the perturbative level.

We have clarified features of the collapse and criterions for its
suppression by the example of the fall to the centre.To prevent the collapse
it is necessary and sufficient that the energy spectrum be bounded from
below.

Noteworthy also is that the energy in supersymmetric theories is always 
formally positive since the lowest eigenvalue of any super\-symmetric 
Hamiltonian is zero. A super\-symmetric system is certain to be immune from 
the collapse if the ground state of this system is {\it stable}. On the other 
hand, the supersymmetry allows to
cancel divergences and sometimes renders the theory finite. One can try to
connect these facts assuming that the supersymmetry shifts the scene to a
realm of lesser dimension where the tendency to collapse is reduced. 
The so-called theorems of non-renormalization in
supersymmetric theories implicitly confirm this assumption. Indeed, the
vanishing of quantum corrections to coupling constants in supersymmetric
theories implies that the picture is semiclassical (the vacuum polarization
is weak) which is feasible only for sufficiently low dimension.

Changing the requirement of the renormalizability to that of the
suppressibility of collapse is a step towards developing a kind of 
`QFT intuition'. L. D. Faddeev \cite{Faddeev} is emphatic that this task is
rather urgent. It is not unlikely that this step will bring us nearer
answering the question: ``Why are three fundamental interactions
renormalizable, while the rest one nonrenormalizable?''.

However, what is the {\it practical} utiliy of this change? The intuitive
understanding of QFT allows to avoid many puzzles without resort to
intricate calculations. For example, it is widely believed that the
divergence of the self-energy in classical electrodynamics $\delta m$ is 
stronger than that in QED $\Sigma $. This belief stems from the fact 
that the divergence of 
$\delta m$ is linear for $D=4$ while the respective divergence of $\Sigma $
is logarithmic. This gave rise to speculations implying that the
quantization can improve the ultraviolet behavior of the theory. But this is
false. Firstly, properties of a classical spinless particle must be compared
with the appropriate properties of the scalar (not spinor) quantum field,
in which case the divergence of $\Sigma $ is quadratic, whence it follows
that the quantization only deteriorates the ultraviolet situation. Secondly,
any $D+1$-dimensional classical field theory relates holographically to a 
$D$-dimensional QFT, and, as was shown in Sec.\ref{irreversib}, $\delta m$
and $\Sigma $ diverge uniformly in such dual theories.

Another long-standing myth was that the introduction of a lattice instead
of the spacetime continuum automatically eliminates all the ultraviolet
divergences. This myth was shattered in 1988 by the following
counter-example \cite{Trivedi}. Let derivatives of the Dirac and scalar
fields in ${\cal L}_{0}$ be replaced by finite differences, and ${\cal L}_{I}
$ be given as ${\cal L}_{I}=-\lambda \phi ^{3}+ig{\bar{\psi}}\psi \phi $,
where arguments of all the fields are taken coincident. One-loop diagrams of
such a lattice model turn out to be divergent, and the model itself is
nonrenormalizable. Lattice theories can suffer from their own ultraviolet
infinities!

Armed with the idea of the kinetic dominance, this conclusion would not have
appeared so much sensational. Recall that any lattice theory is a special
instance of  continual theories with nonlocal form-factors. Indeed,
given the finite differences $\chi (x+l\,{\hat{n}})-\chi (x)$, the free
Lagrangian ${\cal L}_{0}$ is smeared out by the nonlocal operators of the
form $\exp (l\,{\hat{n}}\cdot \partial )-1$. Such a smearing is spread over
a finite region of size $l$. Then the estimate $O(\epsilon ^{-2D})$ in (\ref
{L-0}) is substituted by $O(l^{-2D})$. But all the fields in ${\cal L}_{I}$
are localized in the same points, hence the singularity of the expression (%
\ref{L-I}) remains intact. In other words, we are dealing with the model
where singularities of the kinetic term have been smeared out while those of
the interaction term have been preserved. Now the condition (\ref{crit-reg})
does not hold, the collapse is inevitable, and the model proves
nonrenormalizable.

A fundamental difference between renormalizable and nonrenormalizable
theories is that they correspond to quite different topological pictures of
evolution. The topological manifestation of the nonrenormalizability is
that the mere presence of some special variant of evolution excludes the
possibility of existence of the reverse variant. Renormalizable theories
are free of such a skewness: The reversibility is saved at the cost of
the depletion of topological pictures where all the special variants 
of evolution are absent altogether. However, the question as to 
whether this depletion can result in that every four-dimensional 
renormalizable theory is trivial (maybe in some nonperturbative 
sense) still remains to be solved.

\vskip5mm

\noindent {\Large {\bf Acknowledgment}}

\vskip3mm 
\noindent
The issues touched in this paper and those closely related to
them were discussed with many people at different times. Most beneficial
were old conversations with Ya. B. Zel'dovich and A. Yu. Morozov, and
comparatively recent conversations with R. Haag, D. V. Shirkov, and G. V.
Efimov. I am deeply indebted to them. This work was supported in part by the
International Science and Technology Center under the Project \ \# 840.

\end{document}